\documentclass[twocolumn]{aastex62}

\usepackage{amsmath}
\usepackage{enumitem,booktabs}

\usepackage[referable]{threeparttablex}
\renewlist{tablenotes}{enumerate}{1}
\makeatletter
\setlist[tablenotes]{label=\tnote{\alph*},ref=\alph*,itemsep=\z@,topsep=\z@skip,partopsep=\z@skip,parsep=\z@,itemindent=\z@,labelindent=\tabcolsep,labelsep=.2em,leftmargin=*,align=left,before={\footnotesize}}
\makeatother
\shorttitle{Rotation on wave mixing}
\shortauthors{Varghese et al.}
\begin{document}

\title{Effect of Rotation on Wave Mixing in Intermediate Mass Stars}

\author{A. Varghese}
\affiliation{Newcastle University \\
Newcastle Upon Tyne, United Kingdom}
\affiliation{Universit\'e Paris-Saclay, Universit\'e de Paris, Sorbonne Paris Cit\'e, \\
CEA, CNRS, AIM, 91191 Gif-sur-Yvette, France}
% \email{greg.schwarz@aas.org, gus.muench@aas.org}

\author{R. P. Ratnasingam}
\affiliation{Newcastle University \\
Newcastle Upon Tyne, United Kingdom}

\author{R. Vanon}
\affiliation{ University of Leeds, \\
Leeds, United Kingdom}

\author{P. V. F. Edelmann}
\affiliation{Los Alamos National Laboratory \\
Los Alamos, NM, USA}

\author{S. Mathis}
\affiliation{Universit\'e Paris-Saclay, Universit\'e de Paris, Sorbonne Paris Cit\'e, \\
CEA, CNRS, AIM, 91191 Gif-sur-Yvette, France}

\author{T. M. Rogers}
\affiliation{Newcastle University \\
Newcastle Upon Tyne, United Kingdom}
\affiliation{Planetary Science Institute, Tucson, AZ, USA}



\begin{abstract}

Internal gravity waves (IGWs) are likely to cause mixing in stellar interiors. Studies show that the mixing by these waves changes drastically across age and mass \citep{varghese_2023}. %{\color{red} reference your paper here}. 
Here, we study the effect of rotation on this wave mixing by considering a 7 M$_{\odot}$ model  at ZAMS and midMS. We compare the mixing profiles at a range of rotation rates ($1\times 10^{-5}$, $2\times 10^{-5}$, $3\times 10^{-5}$, $4\times 10^{-5}$ and $1\times 10^{-4}$ rad.s$^{-1}$) and observe that the mixing decreases with decreasing Rossby number.  { {This can be attributed to the effect of rotation on convection which influences the amplitude with which the waves are excited near the convective-radiative interface.}} % enhanced %{\color{red} radiative?, what kind of damping} %
%thermal damping experienced by the waves with rotation. We also determined that in the presence of rotation lower frequency waves are damped more and hence contribute less to mixing. %frequency of the dominant waves contributing to these mixing profiles decreases {\color{red} do you mean the dominant wave frequency decreases?} with rotation suggesting that there are fewer low frequency waves contributing to mixing {\color{red} dont follow}.

\end{abstract}



\section{Introduction} \label{sec:intro}

Internal Gravity Waves (IGWs) are waves propagating in stably stratified fluids with gravity as the restoring force. These are the waves responsible for the Quasi-Biennial Oscillation (QBO) 
%\sout {oscillations} 
\citep{badwin} in Earth's atmosphere and for the turbulent mixing in oceans \citep{munk}. In stars, these waves are stochastically excited either by convective plumes penetrating into the stable stratified region \citep{hurl,mons_2000,pincon2016} or from the  Reynolds stress in the bulk of the convection zone \citep{Kumar,belkam_2009,samadi_2010, LQ}. The amplitudes of these waves are affected by a number of factors such as the density stratification, Brunt-V\"{a}is\"{a}l\"{a} frequency, thermal damping and  geometric effects \citep{rathish2019}. 

Studies suggest that along with other instabilities, IGWs can transport angular momentum and cause mixing in the radiative regions of stars. IGWs were proposed to solve the solar neutrino discrepancy \citep{Press}, %  {NEED REFERENCE HERE}
 explain the Li gap in F type stars and the observed Li abundance in  { {the}} Sun \citep{ gs1991, mon94, ms96}. \cite{Rogers2017} showed that the mixing by IGWs could be treated as a diffusive process %\sout{and is}
 with an amplitude that is proportional to the square of wave amplitudes in the radiation zone. Following their work, \cite{varghese_2023} extended the analysis to stars of different masses and ages. They showed that the mixing is stronger in more massive stars and %\sout{varies drastically} 
decreases as stars age.
\begin{figure*}
\includegraphics[width=1\textwidth]{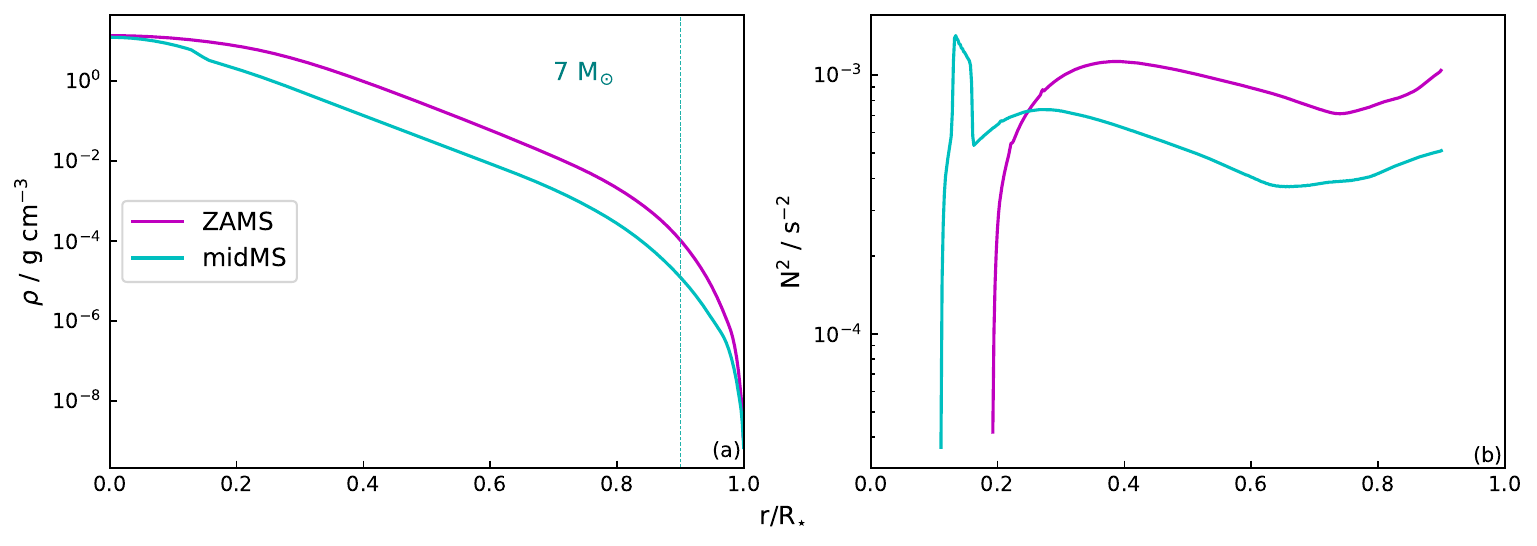}
\caption{ Density  and Brunt-V\"{a}is\"{a}l\"{a} frequency as a function of fractional radius for the 7 M$_{\odot}$  at ZAMS (purple) and midMS (cyan). The vertical dashed line in the density plot indicates the cut off radius of the simulation domain.}
\label{den}
\end{figure*}
%Convective excitation of IGWs can be influenced by a number of factors such as the rotation and Magnetic fields. 

Recent studies demonstrated that the excitation of IGWs is strongly influenced by rotation. The action of the Coriolis acceleration on these waves modifies the damping and the {spatial structure} of these waves resulting in the appearance of new types of waves \citep{mathis2008, mathis2014}. One among these are the gravito - inertial waves (GIWs) where the lower frequencies are significantly affected under the restoring action of buoyancy and Coriolis force \citep{berth, LS}. These waves are excited near the convective - radiative interface, similar to IGWs, either through small scale eddies or by plumes with the coupling between the waves and the turbulence strongly influenced by the Coriolis acceleration \citep{mathis2014}. GIWs are strongly coupled with the turbulence when they are in  the sub-inertial regime ($\omega$ $<$ 2$\Omega$)  remaining as propagative inertial waves in the the convection zone, and are weakly coupled in the super-inertial regime ($\omega$ $>$ 2$\Omega$) when they  { {become}} evanescent \citep{mathis2008,mathis2009} in the convection zone. Rotation, therefore, modifies the coupling between the waves and the  turbulent convective flows subsequently influencing the excitation of gravity waves and GIWs in the stellar radiative interiors \citep{mathis2014}. %is deeply modified by the action of Coriolis acceleration where it is coupled with evanescent super-inertial GIWs or sub-inertial GIWs depending on whether the turbulence is strongly or weakly influenced by rotation \citep{mathis2014}.  These waves experience higher damping with rotation and %The waves modified {\color{red} how are they modified} under rotation 
%are now been driven both by Coriolis acceleration and density stratification \citep{mathis2008} {this sentence doesnt make sense}. 
% These waves are damped more with rotation thereby significantly affecting the angular momentum transport since these are now deposited closer to the generation region \citep{mathis2008, pa_t_c_2007}. 

%  {I think the previous paragraph is meant do describe how waves are influenced by rotation and how this also affects angular momentum transoport and(?) mixing??? needs some attention}

With asteroseismology probing the internal dynamics of stars \citep{timothy2015,timothy2016,papi,szew, sylvian}, gravito - inertial modes \citep{mathis2009, Din2000, ballot2010} have been detected \citep{neiner2012, neiner2020} leading to a new path for exploring stellar interiors. {Excitation of these modes is considered to be  a possible explanation of the variability of certain rapidly rotating stars \citep{balona, drv_1999} as the pure gravity modes and rotation alone couldn't explain the observations \citep{dorbruns, ozz_2019, DBdor}.} %. {\color{red} previous sentence needs more explanation} 
Detection of these modes illustrates the need to include Coriolis acceleration in stellar modelling \citep{may_nature, joey, aerts_2021} and provides asteroseismic calibrations of various parameters such as the overshoot parameter. Gravito-inertial asteroseismology  can give us  information related to convection, buoyancy and rotation \citep{aerts_2021} and therefore, is expected to provide better constraints for numerical simulations.

 The influence of rotation along with IGWs on mixing has been studied previously by \cite{TC5} and \cite{CT_2007}  where they explained the surface Li abundance in low mass stars.  { {They computed 1D stellar models by incorporating the angular momentum transport by meriodional circulation and shear turbulence  following the diffusive/advective formalism \citep{zahn_92, chab_zan, MZ_98} and IGWs. Their model considered the wave excitation  by stochastic eddies similar to the work of \cite{Kumar}. They showed that the shear layer developed near the convection zone due to the deposition of angular momentum acts as a wave filter \citep{GM_filter} and replaced the excited wave spectrum together with the action of shear layer oscillation by a filtered spectrum and diffusion coefficient. %Their studies were successful in solving the long standing problems such as the  surface lithium abundances in low mass stars. 
Inclusion of the angular momentum transport by waves along with the meridional circulations and rotational instabilities in the 1D models by these studies suggested a decrease in the mixing compared with that of rotation alone. These studies have taken into account the indirect influence of the angular momentum transport on the chemical mixing, meaning that IGWs modify the shear that modifies the chemical mixing. Our work focus solely on the direct chemical mixing induced by IGWs in the stellar interiors %. While \cite{Rogers2017} and \cite{varghese_2023} study the transport of chemicals by IGWs in the stellar interior 
considering stars with a convective core and radiative envelope without taking into account the role of angular momentum transport by these waves and the subsequent indirect chemical mixing induced.}} %\cite{R2013} found that the wave amplitudes are larger in rapidly rotating stars. 
 %In this paper we focus on how the mixing by IGWs are influenced by rotation.
 
 Following the work of \cite{Rogers2017} and \cite{varghese_2023}, we study the effect of rotation on wave mixing by considering a 7 M$_{\odot}$ model at ZAMS and midMS. We do not include TAMS models because, as described in \cite{Rathish2020} and \cite{varghese_2023}, these waves are highly attenuated. We achieve this by running 2D simulations with different rotation rates ($\Omega$) using a background reference model from Modules of Experiments in Stellar Astrophysics (MESA) \citep{mesa1, mesa2, mesa3, mesa4, mesa5}. Section \ref{sec:style} introduces the numerical techniques used for obtaining the mixing profiles. Section \ref{sec:floats} presents our findings and Section \ref{con} discuss our conclusions.% {\color{red} need to say now or later that the MESA models themselves are not rotating}
% !!!!!!!!!!!!!!!!!!!!!!!!!!!!Talon and Charbonnel (2007) CHECK LATER!!!!!!!!!!!!!!!!!!!!!!!!!!!!!!!%%%%
\section{Background and Numerical Setup} \label{sec:style}

\subsection{Two-Dimensional Hydrodynamical and Tracer Particle simulations}%\rp{Two-Dimensional Hydrodynamical}}
% \rp{You are starting this section with "We obtained the background reference models" without explaining what "the" refers to here. I would put an introductory sentence to just say that we did 2D simulations and coupled it with tracer particle simulations. Then, say that we used mesa to generate the reference state values. }
% \rp{You have elaborated more on the tracer particle simulation in this section but not much on the 2D sims. I understand that you worked more on the tracer particle code but you did run the 2D sims for this work yourself yes? I think you should put in the equations and elaborate more on the 2D sims here.}
We conducted 2D simulations to study IGWs in stellar interiors and coupled it with tracer particle simulations to determine the mixing by these waves. We generated the background reference models from MESA for a non - rotating 7 M$_\odot$ star at zero age  main sequence (ZAMS, core hydrogen Mass fraction, Xc=$0.70$) and at mid-main sequence (midMS, Xc=$0.35$)  { {with the metalicity set to Z$=0.02$.}} %{\color{red} rotating?}%
We set all stellar parameters similar to those in \cite{rathish_2023} and \cite{varghese_2023}.  { {The inlists used to generate the models are  available in \href{https://zenodo.org/record/2596370\#.Yn5quDnMJUR}{zenodo}.}} The simulations solve the Navier - Stokes equations in the anelastic approximation by considering an equatorial slice of the star with stress-free, isothermal and impermeable boundary conditions similar to that of \cite{R2013} and \cite{rathish_2023}. % \rp{Unneccessary sentence. Just put the references in the sentence before, or say something like ", similar to ..."}.

 The equations are given by

\begin{equation}\label{cont_a}
\nabla.(\bar{\rho}\Vec{v}) = 0,
\end{equation}
\begin{align}\label{mom_a}
\frac{\partial \Vec{v}}{\partial t} + (\Vec{v}.\nabla)\Vec{v}  & = - \nabla \left(\frac{P}{\bar{\rho}}\right) -C\bar{g}\hat{r} + 2(\Vec{v}\times\Vec{\Omega}) \\ 
 & + \bar{\nu}\left(\nabla^{2}v + \frac{1}{3}\nabla(\nabla.v)\right), \nonumber
\end{align}

\begin{align}\label{energy_a}
\frac{\partial T}{\partial t} +  (\Vec{v}.\nabla)T & =-v_{r}\left(\frac{\partial\bar{T}}{\partial r} - (\gamma -1)\bar{T}h_{\rho}\right) \\ \nonumber
& +(\gamma -1)Th_{\rho}v_{r} + \frac{1}{c_v \bar{\rho}} \nabla. \left(c_p\bar{k}\bar{\rho}\nabla T\right) \\
& + \frac{1}{c_v \bar{\rho}} \nabla. \left(c_p\bar{k}\bar{\rho}\nabla \bar{T}\right), \nonumber 
\end{align}   
 where $\bar{\rho}$ and $\bar{k}$ are the reference state density and thermal diffusivity respectively. 
  
 Eqn. \ref{cont_a} is the mass conservation equation  in the anelastic approximation with $\Vec{v}$ as the fluid velocity, while
 Eqn. \ref{mom_a} represents the momentum conservation equation with $\Vec{\Omega}$ as the rotation rate and $\bar{\nu}$ as the kinematic viscosity. $P$ is the reduced pressure defined by \cite{brag} as
 \begin{align}
P = \frac{p}{\bar{\rho}} + U,
 \end{align}
 where $U$ is the gravitational potential perturbation  { {which is neglected in our simulations \citep{cow}}} and $p$ is pressure perturbation.
The co-density perturbation $C$ as defined in \cite{Rogers2005} is
\begin{align}
C = -\frac{1}{\bar{T}}\left(T + \frac{1}{g\bar{\rho}}\frac{d\bar{T}}{dz}p\right).
\end{align} 

Eqn \ref{energy_a} is the energy conservation equation written as a  temperature equation where  $v_{r}$ is the radial velocity and $c_{v}$ is the specific heat capacity at constant volume. The first term in the right hand side of Eqn. \ref{energy_a} represents the super - or subadiabaticity \citep{R2013} which drives the convection in our simulation. It is set to a constant positive value in the convection zone and is calculated from the 1D MESA model in the radiation zone. %The temperature perturbation T is set to zero at the bottom and top boundaries   in our simulation.% \textcolor{purple}{The larger value of the thermal diffusivity used in the simulation enhances the stellar flux and therefore, the amplitude of the Gaussian function  $\bar{Q}$, is determined such that this enhanced thermal diffusivity is compensated by the integrated flux through the system.} {\color{red}I DONT THINK THIS IS CORRECT}

Eqns. \ref{cont_a}-\ref{energy_a}  are  solved  using  a  Fourier  decomposition  method  in  the  azimuthal direction ($\theta$) and a finite difference scheme on a non-uniform grid in the radial direction ($r$). The variables are updated using the Adam-Bashforth explicit method  for the non-linear terms and Crank-Nicolson implicit time - stepping method for the linear terms. %The radial and horizontal velocities ($v_{r}$ and $v_{\theta}$) are calculated from the stream function, $\psi$ through the relations 

% \begin{equation}
%  v_{r} = \frac{1}{\rho r}\frac{\partial \psi}{\partial \theta} \hspace{0.5cm} \text{and} \hspace{0.5cm}  v_{\theta} = -\frac{1}{\rho}\frac{\partial \psi}{\partial r}.
%  \end{equation}

To maintain numerical stability, the simulation domain is cut-off at 90 $\%$ of the total stellar radius, where the density drops beyond $\sim$ 6 orders 
%{\color{red}orders?
of magnitude from the stellar center as %. The background density and Brunt-V\"{a}is\"{a}l\"{a} frequency profiles are 
shown in Fig. \ref{den}.
%\rp{maybe join this with previous sentence by adding "(see Fig. 1)"?}.
We also maintain constant thermal ($\kappa$) and viscous ($\nu$) diffusivities at $5\times 10^{12}$ $\mathrm{cm^{2}s^{-1}}$ in all our simulations  { {and thus the Prandtl number, $Pr$,
\begin{equation}
    Pr = \frac{\nu}{\kappa},
\end{equation}
is equal to $1$ throughout the simulation domain. We have considered this value which is much higher than the actual value in the stellar interior to ensure numerical stability. Recent work by \cite{vanon_2023} could achieve a mild improvement on the $Pr$ ( $Pr$ $\sim$ $5–15$ in the convection zone and $\sim$ $0.02–0.4$ in the radiative envelope) in their three dimensional simulations (see also \cite{sylvian}).}}% to ensure numerical stability\rp{I would remove "to ensure numerical stability" here, as you are repeating this. Just saying what the values are should be enough.}

% \begin{table}[ht]
%   \centering
% % \begin{center}

% % \begin{adjustbox}{width=0.5\textwidth}
% \begin{tblr}{ccc} 
%  \hline
% \SetCell[r=2]{c} $\Omega$ / rad.s$^{-1}$
% & \SetCell[c=2]{c} Convective turnovers & \\
% \hline
% & ZAMS & midMS & & \\
%  \hline
% 1$\times 10^{-5}$ &  97 & 133 \\
% % \hline
% 2$\times 10^{-5}$  & 107 & 119\\
% % \hline
% 3$\times 10^{-5}$ & 93 & 110 \\
% % \hline
% 4$\times 10^{-5}$  & 103 &74\\
% % \hline
% 1$\times 10^{-4}$  & 115 & 44\\
%  \hline
% \end{tblr}
% % \end{adjustbox}
% % \end{center}
% \caption{The total time interval the simulations are run in terms of convective turnover times for ZAMS and midMS model.}
% % \footnotetext[1]{vrms}
% \label{table_1}
% \end{table}
\begin{table}[ht]
  \centering
  \begin{threeparttable}
    \begin{tabular}{lcccc}
      \toprule
      & \multicolumn{2}{c}{Convective turnovers}\\
      \cmidrule(lr){2-3}
      $\Omega$ / rad.s$^{-1}$   &   ZAMS &    midMS   \\
      \midrule
      1$\times 10^{-5}$ &  97 & 133 \\
      2$\times 10^{-5}$  & 107 & 119\\
      3$\times 10^{-5}$ & 93 & 110 \\
      4$\times 10^{-5}$  & 103 &74\\
      1$\times 10^{-4}$  & 115 & 44 \tnotex{tnote:robots-r1}\\
      \bottomrule
    \end{tabular}
    \begin{tablenotes}
      \item\label{tnote:robots-r1}The simulation is run until the vrms attained steady state as shown in Fig. \ref{vrms_7zm}.
      % \item\label{tnote:robots-r2}As recommended by \emph{Robot Review}.
      % \item\label{tnote:robots-r3}That is, X-Ray vision, as proposed by \emph{Mechanical Maniacs}.
    \end{tablenotes}
  \end{threeparttable}
  \caption{The total time interval the simulations are run in terms of convective turnover times for ZAMS and midMS model.}
  \label{table_1}
\end{table}

\begin{figure*}
\includegraphics[width=\textwidth]{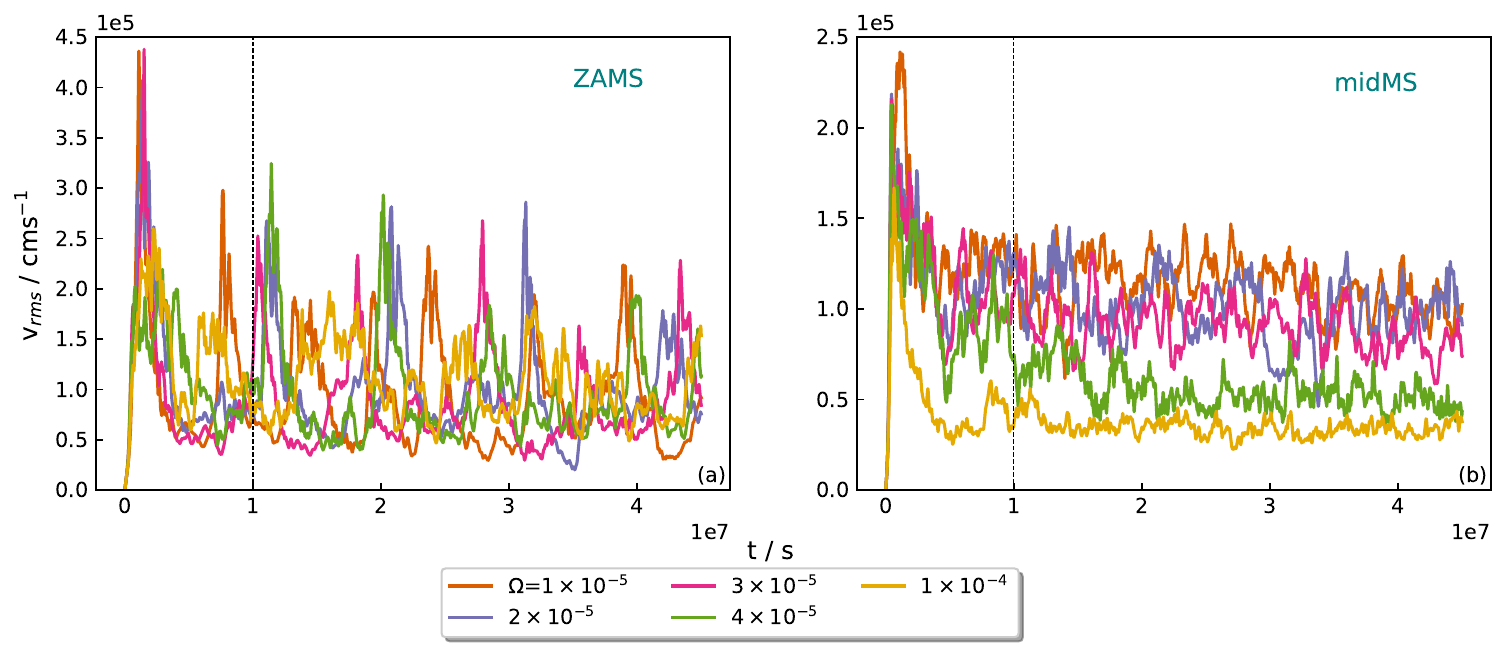}
\caption{v$_{rms}$ averaged over the convection zone for (a) ZAMS, (b) midMS at different angular velocity values.  { {The vertical dashed line indicates the time from which we chose the data for tracer particle simulations.}}}
\label{vrms_7zm}
\end{figure*}
% \footnote{The simulation is run long enough for the vrms to attain the steady state as shown in  Fig. \ref{vrms_7zm} (b)}
\begin{figure*}
\includegraphics[width=1\textwidth]{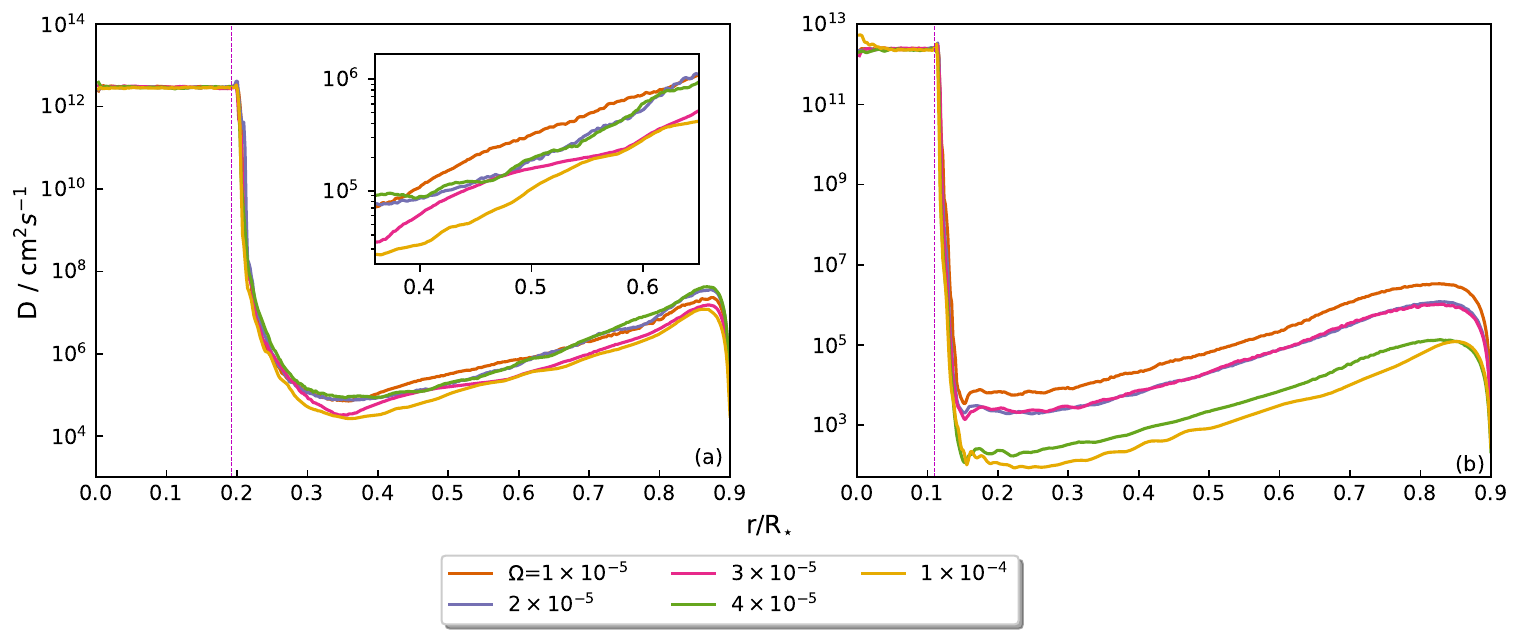}
\caption{Diffusion coefficient as a function of fractional radius for 7 M$_{\odot}$  { {at (a) ZAMS and (b) midMS models}} at $\Omega = 1\times 10^{-5}$ (orange), $2\times 10^{-5}$ (blue), $3\times 10^{-5}$ (pink),  $4\times 10^{-5}$ (light green) and $1\times 10^{-4}$ (yellow) rad.s$^{-1}$. The vertical dashed line represents the convective-radiative interface for each model.}
\label{MP}
\end{figure*}
We then set the initial solid body rotation rate as  { {$1\times 10^{-5}$, $2\times 10^{-5}$, $3\times 10^{-5}$, $4\times 10^{-5}$ and $1\times 10^{-4}$ rad.s$^{-1}$ which in terms of critical rotation rate, v$_{c}$= $\sqrt{GM/R^3}$,  are $0.035\mathrm{v_c}$, $0.07\mathrm{v_c}$, $0.1\mathrm{v_c}$, $0.14\mathrm{v_c}$ and $0.35\mathrm{v_c}$ for the ZAMS models and $0.069\mathrm{v_c}$, $0.13\mathrm{v_c}$, $0.20\mathrm{v_c}$, $0.27\mathrm{v_c}$ and $0.69\mathrm{v_c}$\footnote{This kind of rotation likely distorts the star such that our cylindrical geometry is not appropriate.} for the midMS models, }} %to a range of values from $1 \times 10^{-5}$ %{\color{red} are you keeping zero?}
%to $1\times 10^{-4}$ rad.s$^{-1}$ (up to 35 $\%$ of the critical rotation rate (v$_{c}$= $\sqrt{GM/R^3}$) for ZAMS model and 0.69v$_{c}$\footnote{This kind of rotation likely distorts the star such that our cylindrical geometry is not appropriate.} for the midMS models) %{\color{red} need a footnote here to say that this kind of rotation likely distorts the star such that the geomtry is not entirely appropriate} 
to study the effect of rotation on waves.  We run all the simulations to a total of $4.5\times 10^{7}$s, which is approximately 225 wave crossing times of a $5 \mu$Hz wave (see Table \ref{table_1} for time interval in terms of convective turnover times).  { {Studies by \cite{R2013} determined that there was no significant variations between the models with an initial differential rotation and the ones with an initial solid body rotation. Hence, we also expect the diffusion coefficients calculated in this work to have a very small dependence on whether the model has an initial differential rotation or solid body rotation.}}  %\rp{This sentence is a little convulated, I think. You don't really have to say the wave crossing time because it varies for waves of different frequencies, so I recommend just putting the convective turnover table reference}) 
%{\color{red} should say what this is in terms of both convective turn over time and wave crossing time for a given wave frequency/scale} 
Fig. \ref{vrms_7zm} shows the root mean squared velocity averaged over the convection zone as a function of time
for ZAMS and midMS models. We note that the $v_{rms}$ attained a steady state from $t = 1\times10^{7}$s (indicated as vertical dashed lines in Fig. \ref{vrms_7zm}) and therefore we chose the velocity data saved at a regular time intervals from this value for our further analysis. % but with periodic bursts in ZAMS model. We find these bursts to coincide with the integral multiple of rotational period ($2 \pi/\Omega)$. 
%We then chose the velocity data saved at a regular time interval from $1 \times10^{7}$s %\footnote{We note that the amplitudes of IGWs in our 2D simulation followed a steady--state evolution, that is , it remained constant over time from $1 \times 10^{7}$s} % {from what value??}}
%for our further analysis (see Fig. \ref{vrms_7zm}). % {Again, should show vrms to show steady state reached}

\begin{figure*}
\includegraphics[width=1\textwidth]{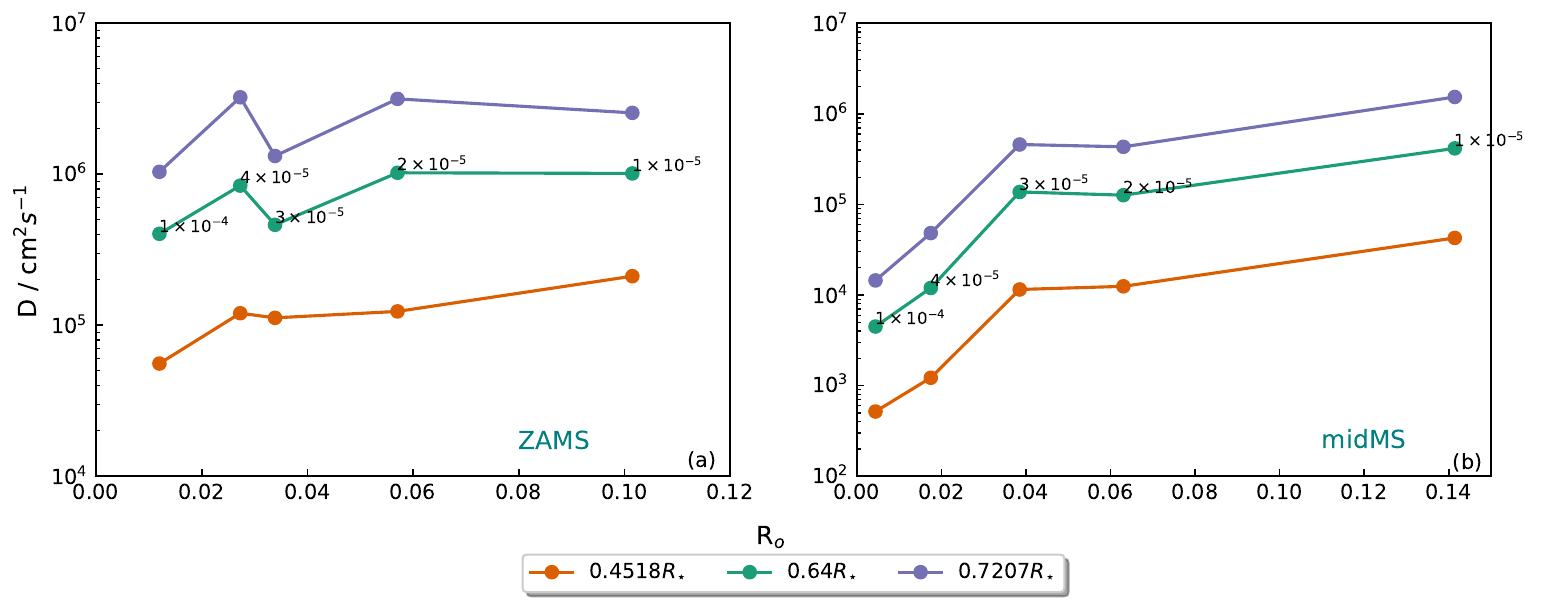}
\caption{Diffusion coefficient as a function of Rossby Number for a 7 M$_{\odot}$ model at (a) ZAMS and (b) midMS for $\Omega = 1\times 10^{-5}$, $2\times 10^{-5}$, $3\times 10^{-5}$, $4\times 10^{-5}$ and $1\times 10^{-4}$ rad.s$^{-1}$ at three different radii.}
\label{ro_mp}
\end{figure*}
 To determine the mixing by IGWs, we introduced $\mathcal{N}$ tracer particles in to our simulations and tracked their trajectories over a time $T$. We calculated the diffusion coefficient, $D$, based on the equations given by \cite{Rogers2017},
 \begin{align} \label{dif}
 D(r,\tau) = \frac{Q(r,\tau)}{2\tau n(r,\tau)}-\frac{P(r,\tau)^{2}}{2\tau n(r,\tau)^{2}},
 \end{align} 
 where $n(r,\tau)$, is the number of sub-trajectories starting at $r$ with a duration of $\tau$, $P(r,\tau)$ is the sum of the lengths of the sub-trajectories and $Q(r,\tau)$ is the sum of the square of these sub-trajectories. More details on the calculation of $n$, $P$, $Q$ and $D(r,\tau)$ can be found in \cite{Rogers2017}.
 
 We plotted the diffusion coefficients as a function of radius at different time differences \footnote{ The smaller $\tau$ has contributions from more time steps compared to a larger value of $\tau$. As an example, consider T$= 100$ with time step $= 1$. Then, for $\tau =5$, it has contributions from $5-0$, $6-1$, $7-2$ and so on, whereas $\tau=97$ can result only from $100-3$, $99-2$ and $98-1$.}, $\tau$, % {The definition of time difference here is a little unclear so it would be good to say what it is physically or relate it to $\tau$.} 
 and chose a time difference ($\tau = 2\times10^{7}$s) such that it has contributions from sufficient number of time steps and the {amplitude of the profiles are converged} (Fig. \ref{MP}). We then used this profile for all our further analysis. %We follow the similar numerical procedure described in \cite{R2013} and rathish 2022.
 %{\color{red} sample figure should be here with explanation of profile BEFORE any theoretical mixing.}

% \section{RESULTS} \label{sec:floats}

\section{RESULTS} \label{sec:floats}
\subsection{Mixing profiles at different rotation}\label{mp_dr}

Fig. \ref{MP} shows the radial diffusion profiles for the 7 M$_{\odot}$ model at ZAMS (left) and midMS (right) for rotation rates, $\Omega = 1\times 10^{-5}$ (orange),  $2\times 10^{-5}$ (blue) $3\times 10^{-5}$ (pink), $4\times 10^{-5}$ (green) and $1\times 10^{-4}$ (yellow) rad.s$^{-1}$. %\rp{This sentence needs to be changed to match the figure}.
We note that all the profiles follow the same trend, that is increasing towards the surface as we move from the convective-radiative interface as expected from \cite{Rogers2017}. This increase can be explained by the change in the wave amplitude due to the balance between decreasing density stratification and the increasing thermal damping towards the stellar surface. %\rp{You can be more specific here, for example "explained by the amplitudes changes experienced by IGWs due to balance between the decreasing density profile and increasing thermal damping towards the stellar surface."}.
However, there are three notable features in the plot which are (i) a particular drop in the amplitude of mixing profile (almost an order of magnitude) as we move from $3\times 10^{-5}$ to $4\times 10^{-5}$ in the midMS model. (ii) The difference between the amount of mixing at slowest rotation and the fastest rotating model is different in ZAMS and midMS. The zoomed plot in Fig. \ref{MP} (a) gives a clearer picture of the trend in ZAMS model. In ZAMS we see a modest change of a factor of approximately 3 while at midMS we see a substantial change of more than an order of magnitude. (iii) We note that the overall  mixing decreases with increasing rotation rate for both ages. 

Before looking in detail at these features, %will be addresses in the following sections. % {what later section??}.%  {\color{red} this is not obvious for the ZAMS models, you need a blow up here}. 
% For better understanding of the variation of mixing with rotation,
we compared %the values {\color{red} what values? 
the diffusion coefficients at a variety of radii as a function of  {the} convective Rossby number, R$_{o}$ %\sout{which is the ratio of inertial forces to Coriolis forces. This is }
defined as,
 {
\begin{equation}
    R_{o}=\frac{v_{rms}}{2\Omega L},
\end{equation}}
where v$_{rms}$ is the averaged root mean squared velocity in the convection zone, $L$ is the extent of the convection zone and $\Omega$ is the rotational frequency. Figure \ref{ro_mp} shows the diffusion coefficients as a function of R$_{o}$  at different radii for all the models studied.  { {We note that there is a clear break in the midMS models at $R_o$ $\sim$ 0.04, below which where we see a steep decline in the diffusion coefficient.}}  %\rp{Try not to use shortened words at the start of a sentence/paragraph. So, Figure 1 instead of Fig. 1 here}.
Regardless of the age of the star or the radius chosen, we see that the trend remains the same, that is, the diffusion coefficient increases with increasing Rossby number.    %{\color{red} quantify change here...In ZAMS we see a modest change of a factor of approximately 2?? while at MIDMS we see a substantial change of more than an order of magnitude, will need to discuss this difference later}
 { {To understand this observed trend; i.e. the wave induced mixing diminishes with rotation, we will study the variation of the damping rate and of the excitation source as a function of rotation in the following sections.}}
%This can be attributed to the extra damping experienced by the waves due to the Coriolis force.  This will be explained further in the following sections. 

\subsection{Linear Theory} 
%{\color{red} WHY IS THEORETICAL MIXING PROFILES IN THE SETUP SECTION??? THIS SHOULDNT BE HERE IT SHOULD BE IN RESULTS}
% Before looking at the influence of rotation on the damping profile  {you've changed order here so this sentence doesnt make sense anymore}, here 

In this section we discuss the theoretical prescription from \cite{Rogers2017}, where they determined the diffusion coefficient $D$,
\begin{align} \label{eq:1}
D =Av_{wave} ^{2} (\omega, l,r) ,
\end{align}
with the coefficient $A \sim$ $1$s  \footnote{The constant A is unknown and likely depends on simulation parameters such as the total time domain considered. We expect that this parameter can be constrained using asteroseismology \citep{may_nature}.} %\rp{The footnote looks like a power of 3}
\citep{Rogers2017, varghese_2023} and the wave amplitude calculated using the linear theory without considering rotation from \cite{rathish2019}, 
%{\color{red} NEED TO NOTE HERE THAT THIS WAS JUST A FIT AND THE VALUE OF A IS UNCERTAIN}
\begin{align} \label{eq:2}
 v_{wave}(\omega,l,r) = v_{0}(\omega,l,r)\left(\frac{\rho}{\rho_{0}}\right)^{-\frac{1}{2}}\left(\frac{r}{r_{0}}\right)^{-1} \\
 \left(\frac{N^{2}-\omega^{2}}{N_{0}^{2}-\omega^{2}}\right)^{-\frac{1}{4}}e^{-\frac{\mathcal{T}}{2}}, \nonumber
\end{align}
where $\rho_0$, $r_0$ and $N_0$ are the density, radius and the Brunt-V\"{a}is\"{a}l\"{a} Frequency at the initial reference point. $v_{0}$($\omega$,l,r) is the initial wave amplitude for  a given frequency and wavenumber.
The damping coefficient, $\mathcal{T}$ as given by \cite{Kumar} and \cite{zan97} is expressed as:
\begin{equation} \label{eq:3}
\hspace{-0.0cm}\mathcal{T} (\omega, l,r) =  \int_{r_{0}}^{r} \frac{16\sigma T^{3}}{3\rho^{2}\kappa c_{p}}\left(\frac{(l(l+1))^{\frac{3}{2}}N^{3}}{r^{3}\omega^{4}}\right)\left(1-\frac{\omega^{2}}{N^{2}}\right)^{\frac{1}{2}}dr.
\end{equation}
where $\sigma$, $T$, $c_p$, $\kappa$ and $l$ are the Stefan-Boltzmann constant, temperature, specific heat capacity at constant pressure, opacity and wavenumber. 

Using these equations \cite{varghese_2023} found that the dominant waves contributing to the mixing profiles are low frequency waves within the range of $4 - 9 \mu$Hz  with $5 \mu$Hz being the dominant wave for the $7$M$_{\odot}$ ZAMS and midMS non-rotating models. As discussed in \cite{rathish2019} these are the waves generated with amplitudes large enough to escape the thermal damping, but low enough that they are efficiently generated by convection. While many frequencies contribute to the mixing profile, the aim of identifying the dominant waves contributing the mixing profiles was to make the inclusion of wave mixing in 1D stellar evolution models easier.
%{\color{red} again, need to put this later and then need more description as to why you would do this}We used the above Eqn. \ref{eq:1}-\ref{eq:3} to determine the dominant frequencies contributing to the mixing profiles obtained from our simulations \citep{varghese_2023} for the non rotating models.

%  {WHY IS FIGURE 4 here?? It should be above to demonstrate steady state but with periodic bursts *maybe* coinciding wiht rotation rate???}
\subsection{Linear Theory including rotation}\label{th_mp}
%{\color{red} AGAIN, WHY IS THIS HERE? THIS ISNT PART OF THE SETUP ITS TOTALLY UNCLEAR AT THIS POINT OF THE PAPPER WHY YOU'RE DOING THIS}
To theoretically determine the effect of rotation on waves, we followed the approach of \cite{Press} and  \cite{Rathish2020} in solving the linearised hydrodynamic equations in the anelastic approximation, but with the inclusion of rotation term in the momentum equation (Eqn. \ref{mom}). The linearised equations neglecting thermal and viscous diffusivities are given below: %\rp{$\Omega$ is a vector}.
\begin{align}\label{cont}
\nabla.(\bar{\rho}v) & = 0,
\end{align}
\begin{align}\label{mom}
\frac{\partial v}{\partial t}   & = - \nabla \left(\frac{P}{\bar{\rho}}\right) -C\bar{g}\hat{r} + 2(v\times\Vec{\Omega}),% - (v.\nabla)v \\ 
%  & + \bar{\nu}\left(\nabla^{2}v + \frac{1}{3}\nabla(\nabla.v)\right), \nonumber
\end{align}
and
\begin{align}\label{energy}
\frac{\partial T}{\partial t}  & =-v_{r}\left(\frac{\partial\bar{T}}{\partial r} - (\gamma -1)\bar{T}h_{\rho}\right). % - (v.\nabla)T\\
% & +(\gamma -1)Th_{\rho}v_{r} +\gamma\bar{K}\left[\nabla^{2}T + h_{\rho}\frac{\partial T}{\partial r}\right] \nonumber \\
% & + \gamma\bar{K}\left[\nabla^{2}\bar{T} + h_{\rho} \frac{\partial\bar{T}}{\partial r}\right] + \frac{\bar{Q}}{c_{\nu}}. \nonumber
\end{align}   
We reduce the above %\rp{It sounds like the above equations have viscous and thermal damping and you are reducing them by neglecting these effects. They have already been neglected so mention this in the sentence before the equations}
to get the following second order differential equation for wave propagation considering a wave ansatz, $v_{r}(r,\theta,z) \propto v_{r}(r)e^{im\theta}e^{-i\omega t}$,
\begin{align}\label{red_LT}
0  = & \frac{\partial^2 \alpha}{\partial r^2} + \left[\frac{N^2}{\omega^2} - 1 \right]\frac{m^2}{r^2}\alpha +\frac{1}{2}\left[\frac{\partial h_{\rho}}{\partial r} - h_{\rho} ^{2}  +\frac{h_{\rho}}{r}\right]\alpha  \nonumber  \\
 & + \frac{1}{4r^2}\alpha - \frac{2\Omega h_{\rho}m}{\omega r}\alpha.
\end{align}

Here, $\alpha= v_{r}\bar{\rho}^{\frac{1}{2}}r^{\frac{3}{2}}$, $m$ is the 2D Fourier basis wavenumber, $\omega$ is the angular frequency and $h_\rho$ is the inverse density scale height. %As already described in \cite{Rathish2020}, the oscillatory term (OT) is given by, 
% \begin{equation}
%     \left[\frac{N^{2}}{\omega^{2}} -1\right] \frac{m^{2}}{r^{2}},
% \end{equation}
% the density term (DT),
% \begin{equation}
%     \frac{1}{2}\left[\frac{\partial h_{\rho}}{\partial r} - \frac{h_{\rho} ^{2}}{2}  +\frac{h_{\rho}}{r}\right] ,
% \end{equation}
% and the geometric term (GT), ${1}/{4r^2}$. The turning point is defined as the radius at which the ratio of the OT to that of the DT is less than 1. Waves lose their wave like behaviour beyond this point, thereby experiencing extra damping. The study also showed that the turning point is located a lower fraction of the stellar radius in older stars compared to that of the younger stars. Based on this, \cite{varghese_2023} had shown the effect of turning point on the mixing profiles across different ages and concluded that the extra damping experienced by the waves near the turning point significantly affects the mixing in stars, particularly in older stars.  
\begin{figure*}
\includegraphics[width=1\textwidth]{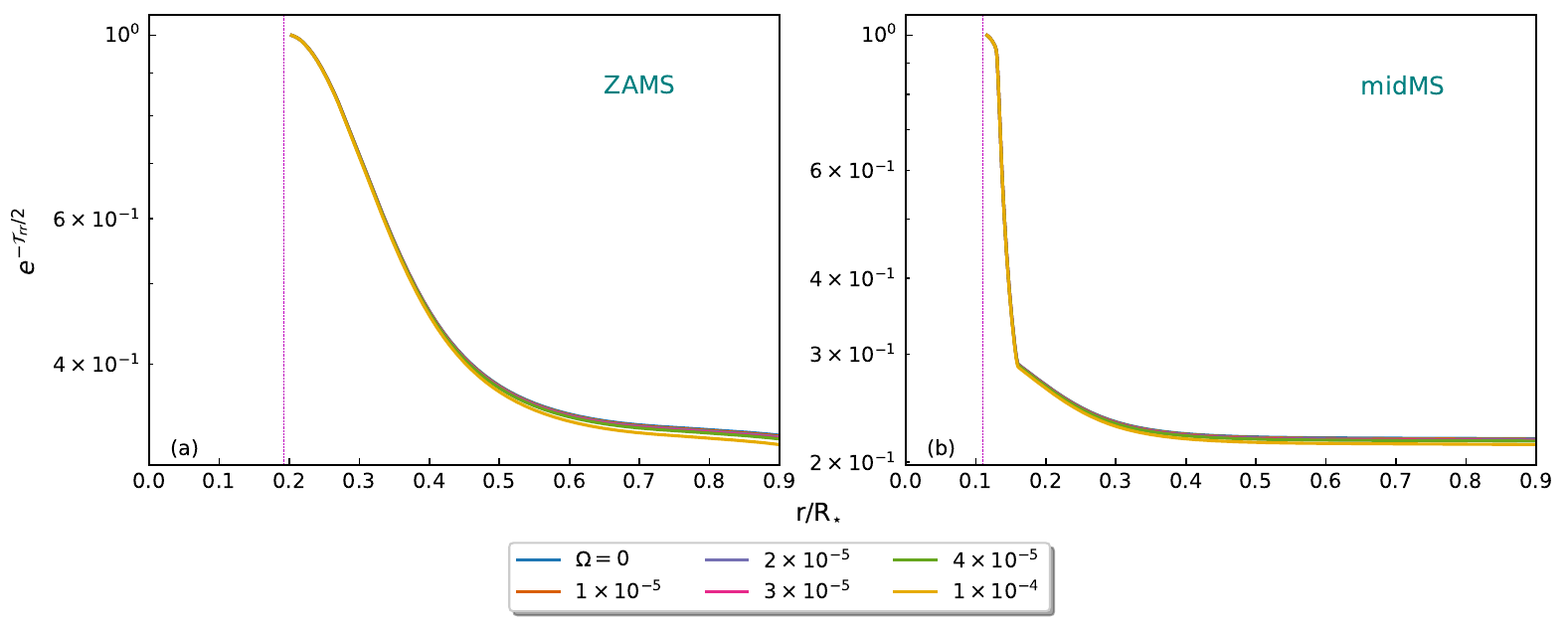}
\caption{The attenuation coefficient ($\exp{-(\mathcal{T}_{rr}/2)}$) as a function of fractional radius for the 7 M$_{\odot}$  at (a) ZAMS and (b) midMS at different angular velocity values considering a $5 \mu$Hz wave at $l = 1$.}
\label{damp_plot}
\end{figure*}
In this work, we introduce the term including the rotational effects, $2\Omega h_{\rho}m/\omega r$. Here we look at the combined effect of radiative damping, density stratification and rotation. %This term reduces the effect of damping causing the turning point to be shifted to higher radius.

\subsection{Radiative damping and rotation}\label{damp}
%{\color{red} SAME AS ABOVE, SHOULDNT BE HERE}

Earlier studies of \cite{Press} and \cite{Kumar} defines the wave damping opacity as 
\begin{equation} \label{damp_ratio}
\hspace{-0.0cm}\mathcal{T} (\omega, l,r) =  \int_{r_{0}}^{r} \frac{\gamma[\omega,l,r]}{\lvert v_{g}[\omega,l,r] \rvert}dr,
\end{equation} 
where the damping rate $\gamma$ is given as

\begin{equation}
    \gamma(\omega,l,r)=Kk_{r}^2,
\end{equation}
with the radial wavenumber $k_{r}^2$,

\begin{equation}\label{rad_wave}
    k_{r}^2=\left(\frac{N^{2}}{\omega^2} - 1 \right )k_{h}^2.
\end{equation}
The vertical group velocity, $v_{g}$ and the thermal diffusivity, K are given as
\begin{align}
    v_{g}(r) = \frac{\partial \omega}{\partial k_{r}} = - \frac{(N^2-\omega^2)^{\frac{1}{2}} \omega^2}{k_{h}N^2},
\end{align}

\begin{equation}
        K(r) = \frac{16\sigma T^3}{3\rho^2\kappa c_p}.
\end{equation}
 In the absence of rotation the horizontal wavenumber, $k_{h}$ is given as
 \begin{equation}
     k_{h}^{2} = \frac{l(l+1)}{r^2}
 \end{equation}
 and the Eqn. \ref{damp_ratio} simplifies to Eqn. \ref{eq:3}.

 % Whereas, in the presence of rotation, the eigenvalues of modes grows, significantly increasing the radial wavenumber (Eqn. \ref{rad_wave}) and therefore the wave damping (Eqn. \ref{damp_ratio}) \citep{pa_t_c_2007}.
 
 From Eqn. \ref{red_LT}, the modified radial wavenumber, k$_{rr}$, in the presence of rotation is given by
 \begin{equation}\label{rot_kr}
     k_{rr}^2= \left(\frac{N^{2}}{\omega^2} - 1 \right )k_{h}^2 - \frac{2\Omega h_{\rho}k_h}{\omega }
 \end{equation}
 { {where $k_h=m/r$. The equation including the effect of rotation on the vertical group velocity with rotation, $v_{g_{rr}}$ is given as:
\begin{equation}
    v_{g_{rr}} = \frac{\partial \omega}{\partial k_{rr}} 
    =-\frac{k_{rr}}{\frac{N^2k_h^2}{\omega^3}-\frac{\Omega h_{\rho}k_h}{\omega^2}}.
\end{equation}
 %The density scale height, $h_{\rho}$ has a negative sign, due to the decreasing density towards the surface and hence, the rotation term increases the radial wavenumber, $k_{rr}$ (and hence, decreases the radial wavelength). 
 Therefore, the damping rate is modified by rotation and is given as $\gamma_{rr}$:
 \begin{equation}
     \gamma_{rr}(\omega,l,r)=Kk_{rr}^2,
 \end{equation}
 thereby influencing the spatial damping opacity in rotating models, $\mathcal{T}_{rr}$
 \begin{align} \label{damp_mod_new}
 \mathcal{T}_{rr} (\omega, l,r) & =  \int_{r_{0}}^{r} \frac{\gamma_{rr}[\omega,l,r]}{\lvert v_{g_{rr}}[\omega,l,r] \rvert}dr, \nonumber \\
\end{align}
with
  \begin{align} \label{damp_mod}
    \mathcal{T}_{rr} (\omega, l,r) = \int_{r_{0}}^{r}\frac{16\sigma T^{3}}{3\rho^{2}\kappa c_{p}} \left(\left[1-\frac{\omega^2}{N^2}\right]k_h^2 - \frac{2\Omega h_{\rho}\omega k_h}{N^2}\right)^{1/2} \\ \nonumber
    \left(\frac{N^3k_h^2}{\omega^4} - \frac{\Omega h_{\rho}N k_h}{\omega^3}\right).
                % -\frac{k_{rr}}{\frac{N^2k_h^2}{\omega^3}-\frac{2\Omega h_{\rho}}{\omega^2r}}
% \hspace{-0.0cm}\mathcal{T}_{rr} (\omega, l,r) =  \int_{r_{0}}^{r} \frac{\gamma_{rr}[\omega,l,r]}{\lvert v_{g}[\omega,l,r] \rvert}dr.
\end{align}} }

 The wave amplitude given by Eqn. \ref{eq:2} can be rewritten as:
 \begin{align} \label{mod_v}
 v_{wave_{rr}}(\omega,l,r) = v_{0}(\omega,l,r)\left(\frac{\rho}{\rho_{0}}\right)^{-\frac{1}{2}}\left(\frac{r}{r_{0}}\right)^{-1} \\
 \left(\frac{(N^{2}-\omega^{2})m^2 - 2\Omega h_{\rho}m\omega r}{(N_{0}^{2}-\omega^{2})m^2 - 2\Omega h_{\rho}m \omega r_{0}}\right)^{-\frac{1}{4}}e^{-\frac{\mathcal{T}_{rr}}{2}}. \nonumber
\end{align}

 { {We now compare the damping experienced by the waves at different rotation rates for the ZAMS and midMS models. This is carried out by plotting the attenuation coefficient ($\exp{-(\mathcal{T}_{rr}/2)}$) calculated with Eqn. \ref{damp_mod} as a function of the fractional radius for a $5 \mu$Hz wave at $l=1$ \footnote{We chose $l=1$ in accordance with \cite{varghese_2023} where they had shown that the higher wavenumbers experience higher thermal damping as expected from Eqn. \ref{eq:3}.} shown in Fig. \ref{damp_plot}. We observe an overall decrease in the attenuation factor with rotation. However, we find that this decrease in the attenuation factor with rotation is not sufficient enough to explain the difference seen in the mixing profiles (Fig. \ref{MP}). Hence, we will now evaluate the influence of rotation on convection and on the subsequent wave excitation in the following sections. We begin by comparing the theoretical diffusion coefficient calculated using Eqns. \ref{damp_mod}, \ref{mod_v} and \ref{eq:1} with the simulation profiles.}}

% We chose $l=1$ in accordance with \cite{varghese_2023} where they had shown that the higher wavenumbers experience higher thermal damping as expected from Eqn. \ref{damp_ratio}.
% Thus, rotation increases the radial wavenumber contributing to a higher damping opacity, $\mathcal{\tau}$ \citep{pa_t_c_2007}. Therefore, the waves experience higher thermal damping with increasing rotation and thus the decrease in the wave mixing seen in Fig. \ref{MP}.

 { {\subsection{Theoretical Mixing Profiles}\label{theory_explain}
We calculated the theoretical diffusion profiles using the Eqns. \ref{damp_mod}, \ref{mod_v} and \ref{eq:1} for each frequencies and wavenumbers. Considering the following relation between the tangential and radial velocity based on the assumption that the frequencies are much smaller than the Brunt-V\"{a}is\"{a}l\"{a} frequency,
\begin{equation}
    \frac{v_{\theta}}{v_r}= \frac{N}{\omega},
\end{equation} 
\begin{figure*}
\includegraphics[width=\textwidth]{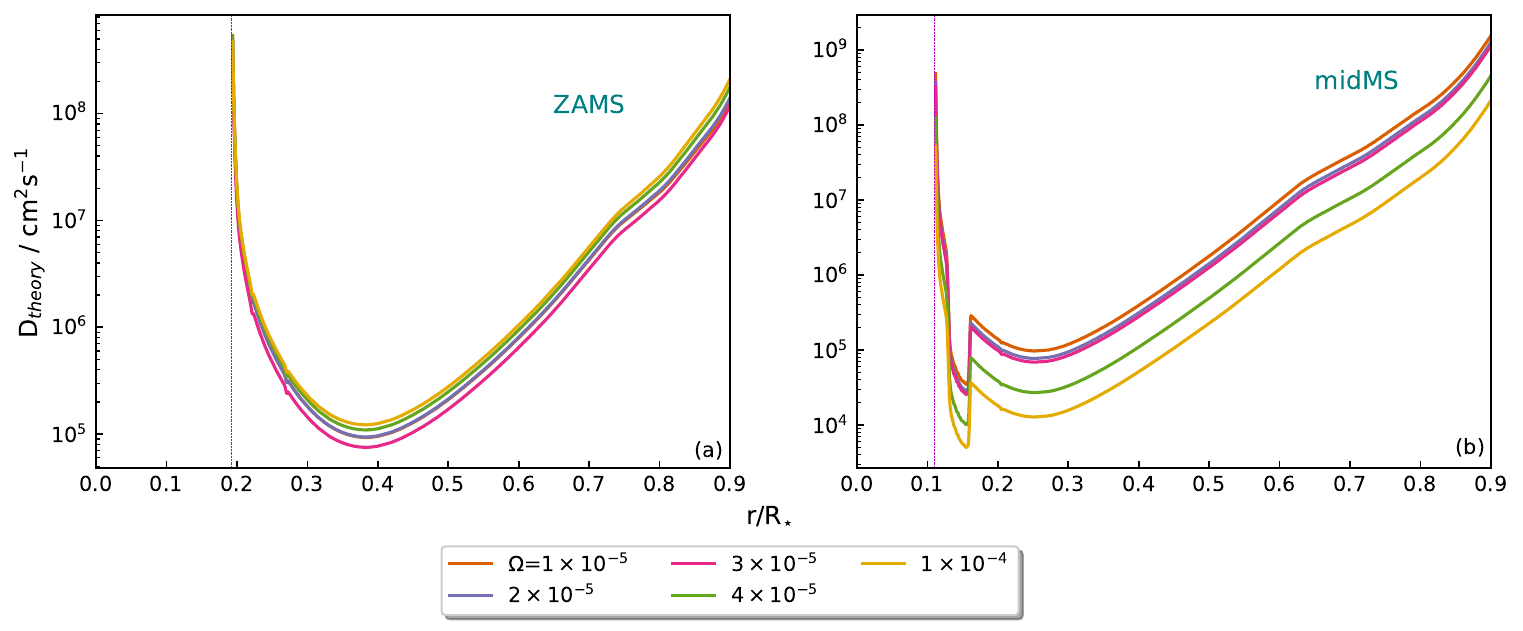}
\caption{Theoretical diffusion coefficient calculated using Eqns. \ref{eq:1}, \ref{damp_mod} and \ref{mod_v} for (a) ZAMS, (b) midMS for different $\Omega$'s as a function of fractional radius for a 5$\mu$Hz wave. The vertical dashed line indicates the convective radiative interface.}
\label{dtheory_mp}
\end{figure*}
we set $v_{0}(\omega,l,r)$  to the root mean squared velocity at a radius just outside the convection zone averaged over the total time domain  multiplied by a factor of $\omega/N$. The wave launch point is chosen such that the Brunt-V\"{a}is\"{a}l\"{a} frequency is greater than $10 \mu$Hz since the dominant waves found from \cite{varghese_2023} were in the range of $4-9$ $\mu$Hz for ZAMS and midMS models at $l=1$. Fig. \ref{dtheory_mp} shows the theoretical profiles calculated for a 5 $\mu$Hz wave for the ZAMS and midMS models. We observe that the theoretical profiles do not follow a trend similar to that of the simulation in the case of ZAMS models. The profiles from the theory follows a similar trend as that of the simulation in midMS models, with a particular drop in the amplitude of the  mixing profiles as we move from $3 \times 10^{-5}$ to $4\times10^{-5}$ rad.s$^{-1}$. However, the amplitude difference is only $\sim$ 2.6 compared to an order of magnitude difference in the simulation. We know that the initial velocity $v_{0}$ is set by the convection, but it is still unclear on how to define this value. Hence, to get a deeper understanding on the choice of the initial velocity and of the difference observed between the different rotation rates, we look in detail at the influence of rotation on convection and the theory from  \cite{august_2020} which also includes further effects of rotation on wave generation and propagation.}}
\subsection{Influence of rotation on convection} \label{rot_con}
Fig. \ref{vrms_7zm} shows an overall decrease in the rms velocity  averaged over the convection zone with rotation. \cite{chandra} demonstrated that the rotation influences the onset of turbulent convection.  { {In addition, \cite{steve} predicted that the mode of convection that carries the more heat is inhibited by the action of rotation building a mixing length theory for rotating convection. \cite{barker_2014} verified this theoretical prediction by conducting high resolution nonlinear numerical simulations in a Cartesian box assuming the Boussinesq approximation. Later, \cite{august_2019} have generalised the initial study by \cite{steve} by taking into account viscous and heat diffusions and have confirmed the results obtained by \cite{steve}. % considered the approach of \cite{steve} further showing that the influence of rotation on convection reduces the convective velocities agreeing with the previous studies. 
Hence the decreases in the convective velocities noted in Fig. \ref{vrms_7zm} can be attributed to the influence of rotation on convection.}}% This influences the amplitude with which the waves are excited near the convective radiative interface and hence the overall mixing as we move further in to the radiation zone.} 

A notable feature in Fig. \ref{MP} is the large decrease in the amplitude of the mixing profile in the radiation zone from $\Omega = 3 \times 10^{-5} $ to $ 4\times 10^{-5} $ rad.s$^{-1}$ in the case of midMS model. We see a decrease of approximately an order of magnitude compared to the other models (approximately a factor of 3 from $\Omega = 1 \times 10^{-5} $ to $2\times 10^{-5} $ rad.s$^{-1}$ and 2 from $\Omega = 4 \times 10^{-5} $ to $ 1\times 10^{-4} $ rad.s$^{-1}$). This can be attributed to the variation of rms velocity in the convection zone. Even though we note an overall decrease in  the rms velocity averaged over the convection zone with rotation, this decrease is very evident for $\Omega = 4 \times 10^{-5} $ at midMS as seen from Fig. \ref{vrms_7zm}.  Under the action of rotation, the critical Rayleigh number (Ra$_{c}$) is given as, 
\begin{equation}
    Ra_c \propto Ek^{-\frac{4}{3}}
\end{equation}
where, the Ekman number, $Ek$ is given as,
\begin{equation}\label{ek}
    Ek=\frac{\nu}{2\Omega L}.
\end{equation}
Hence, $Ra_c$ increases with rotation implying that at higher rotation, stronger forcing is necessary for driving the convection to be in the supercritical regime compared to that of a slowly rotating model \citep{chandra, rrb}. This can also be noted from the values of Reynolds number, $Re$ given in Table \ref{table_rot_re}. Here $Re$ is defined as,
 \begin{equation}
    \mathrm{Re} =  \frac{v_{rms}L}{\nu},
 \end{equation}
\begin{table}[ht]
  \centering
% \begin{center}
% \begin{adjustbox}{\textwidth}
\begin{tabular}{cccc} 
 \hline
 Model & $\Omega$ / rad.s$^{-1}$ &$Re$ & average $v_{rms}$ / cms$^{-1}$  \\ [0.5ex] 
 \hline
  7M$_{\odot}$ ZAMS&  $1\times10^{-5}$&813&93988 \\
% \hline
 & $2\times10^{-5}$&891&103048 \\
 &$3\times10^{-5}$&779&90049 \\
% \hline
% \hline
 & $4\times10^{-5}$&862&99642 \\
% \hline
 & $1\times10^{-4}$&956&110548 \\
 \hline
7M$_{\odot}$ midMS &$1\times10^{-5}$&921&116810 \\
&$2\times10^{-5}$&822&104189 \\

 &$3\times10^{-5}$&764&96811 \\
 &$4\times10^{-5}$&518&65671 \\
 &$1\times10^{-4}$&306&38792 \\
 \hline
\end{tabular}
% \end{adjustbox}
% \end{center}
\caption{Reynolds' number, $Re$ and the averaged $v_{rms}$ in the convection zone for all the models studied.}
\label{table_rot_re}
\end{table}
 where we recall that $L$ is the radial extend of the convection zone. We see that $Re$ remains within $750-950$ for the ZAMS model, whereas, there is a larger variation, $30\%$  decrease between $\Omega = 3 \times 10^{-5} $ to $ 4\times 10^{-5} $ rad.s$^{-1}$, compared to $7\%$ decrease between $\Omega = 2 \times 10^{-5} $ to $ 3\times 10^{-5} $ rad.s$^{-1}$ and $10\%$ decrease between $\Omega = 1 \times 10^{-5} $ to $ 2\times 10^{-5} $ rad.s$^{-1}$  as we move from the slow rotating to fast rotating model in midMS. %This decline in the $Re$ from $\Omega = 3 \times 10^{-5} $ to $ 4\times 10^{-5} $ rad.s$^{-1}$ for the midMS further support our argument that the onset of convection is strongly influenced by rotation. 

%     \begin{figure*}
% \includegraphics[width=\textwidth]{Dflux_7zm.pdf}
% \caption{Diffusion coefficient calculated from the interfacial flux for a $5 \mu$Hz wave for (a) ZAMS, (b) midMS for different $\Omega$'s as a function of fractional radii. The vertical dashed line indicates the convective-radiative interface.}
% \label{dflux_mz}
% \end{figure*}
We therefore infer from the above that the difference in the rms velocity observed is due to the convection being inhibited at higher rotation.  Consequently, higher rotation rates influence the convection significantly such that the waves are generated with a lower amplitude \citep{tak} which in turn explains a decrease in the mixing by these waves. 

The significant difference observed in the $Re$ values in the midMS models compared to that of the ZAMS  {can thus} also explain the considerable variation in the mixing profiles in midMS as we move from the slow to fast rotating models. We see a larger variation across midMS but not in ZAMS, suggesting that 
%because of the difference in the density stratification in ZAMS and midMS. Studies suggest that %along with rotation, the critical Rayleigh number increases with an increase in the density stratification. Since older stars are more stratified than younger stars, this influences the onset of convection, requiring 
the convection needs to be forced even stronger in midMS to be at an equivalent $Re$. The evident decrease observed in the rms velocities of midMS models (Table. \ref{table_rot_re}) must influence the amplitude with which the waves are generated at the interface  { {and the mixing they trigger. To confirm this, we now consider how the kinetic energy of the turbulent convection is transmitted to IGWs at the convective core boundary as a function of rotation.}} 
\begin{figure*}
\includegraphics[width=\textwidth]{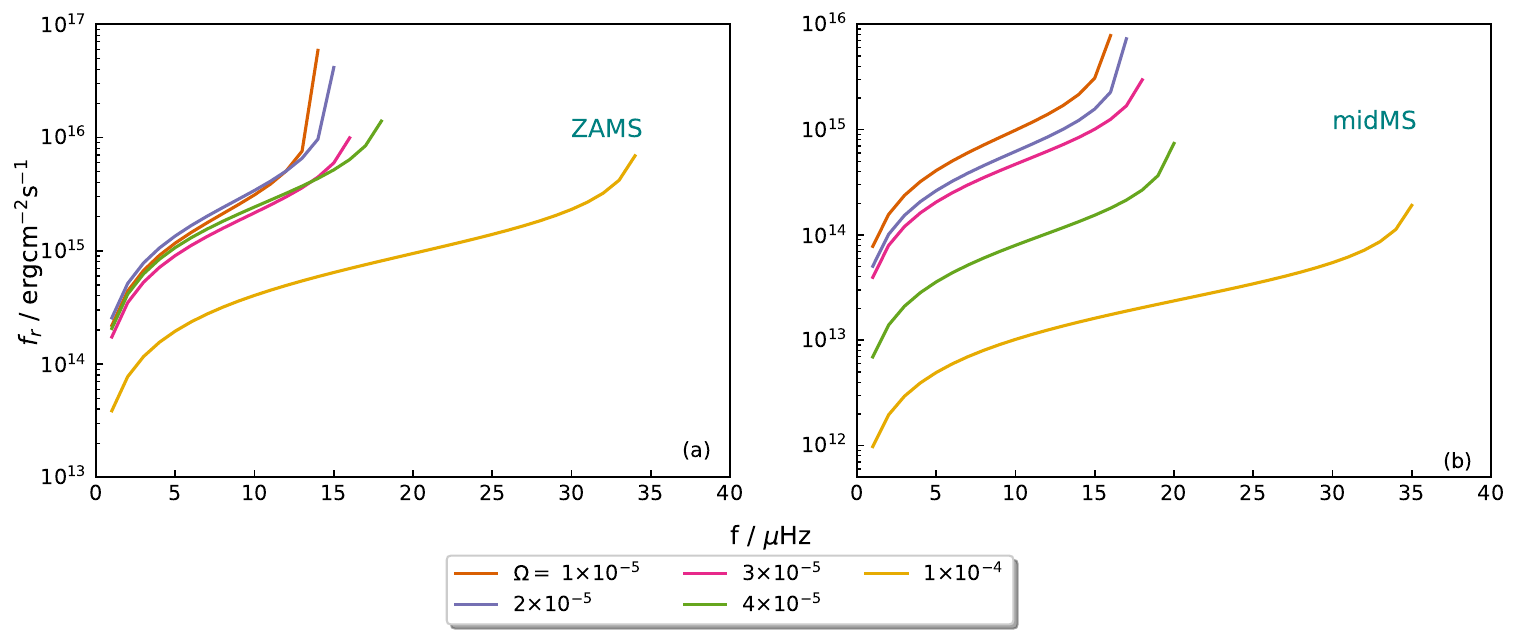}
\caption{Interfacial flux calculated using Eqn. \ref{fz_ag} for (a) ZAMS, (b) midMS for different $\Omega$'s as a function of frequency.}
\label{flux_mz}
\end{figure*}
    \begin{figure*}
\includegraphics[width=\textwidth]{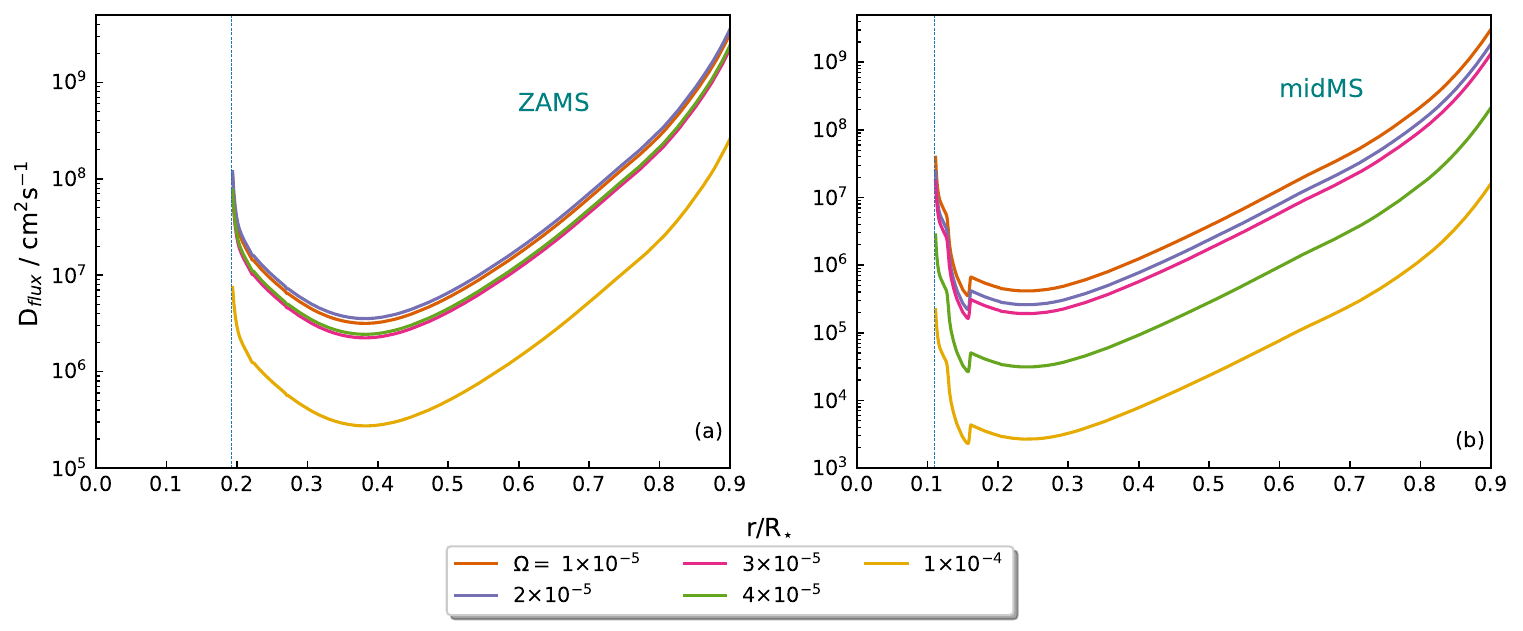}
\caption{Diffusion coefficient calculated from the interfacial flux for a $5 \mu$Hz wave for (a) ZAMS, (b) midMS for different $\Omega$'s as a function of fractional radii. The vertical dashed line indicates the convective-radiative interface.}
\label{dflux_mz}
\end{figure*}

 { {\subsection{Influence of Rotation on Wave Excitation}
 In addition to the effect on convection, rotation affects the wave excitation as shown in \cite{august_2020}.  To evaluate this quantitatively for these simulations, we calculated the flux of the GIWs for all the models studied in this work. We adopted the approach of \cite{august_2020} adapted to our equatorial geometry. The flux was calculated by considering Eqn. $40$ in \cite{august_2020} for wave excitation by pressure fluctuations at the convection/radiation interface and applying it to the peculiar case of the equator. This leads to:
\begin{equation}\label{fz_ag}
       F_r= \frac{1}{2}\rho v_{int}^{3}\frac{\omega}{(N^2 + 4\Omega^2 - \omega^2)^{1/2}}.
\end{equation}
 The above equation relates the wave energy flux in the radial direction to the convective velocity at the convective-radiative interface measured in our simulations, $v_{int}$, which is the root mean squared velocity averaged over the simulation time domain, the wave frequency, $\omega$, and the rotation rate, $\Omega$. Fig. \ref{flux_mz} shows the wave flux as a function of frequencies at each rotation rate for both the ZAMS and midMS models. We observe that the interfacial wave flux decreases with rotation agreeing with the results of \cite{august_2020}. We chose the wave launch point such that the Brunt-V\"{a}is\"{a}l\"{a} frequency is greater than $10 \mu$Hz since the dominant waves found from \cite{varghese_2023} were in the range of $4-9$ $\mu$Hz for ZAMS and midMS models as in section \ref{theory_explain}. The decrease in the wave energy flux seen in Fig. \ref{flux_mz} implies that the transport of chemicals induced by the waves in the stellar interior should decreases with rotation.}} 
 
  { {Our aim now is thus to compare the diffusion coefficients ($D_{flux}$) calculated from the wave flux with that of the simulation. %We therefore determine the wave velocity ($v_w$) from the flux calculated. For this calculation, we use the 
 Considering Eqn. $35$ from \cite{august_2020} adapted to our equatorial geometry, we obtain the general relation between the radial flux of energy and the radial components of GIWs velocity:
 \begin{equation}
        F_r = \frac{\rho}{2k_h}(N^2 + 4 \Omega^2 - \omega^2)^{1/2} v_w^2.
    \end{equation}
    We thus obtain the wave velocity at the interface, $v_w$, as,
    \begin{equation}
        v_w^2 = \frac{2mF_r}{r \rho (N^2 + 4 \Omega^2 - \omega^2)^{1/2} }.
    \end{equation}
    We now substitute $v_w$ as $v_0$ in Eqn. \ref{mod_v} and calculate the diffusion coefficient using Eqn. \ref{eq:1} for each frequency and wave number. Fig. \ref{dflux_mz} shows the calculated diffusion coefficient for a $5\mu$Hz wave. We see that the difference between these theoretical profiles at different rotation rates agrees reasonably well with that of the difference between the simulation mixing profiles shown in Fig. \ref{MP} for models up to rotation rate, $\Omega = 4\times 10^{-5}$ rad.s$^{-1}$. The difference is approximately a factor of 1.6 from $\Omega = 1 \times 10^{-5} $ to $2\times 10^{-5} $ rad.s$^{-1}$, 1.3 from $\Omega = 2 \times 10^{-5} $ to $ 3\times 10^{-5} $ rad.s$^{-1}$ and 6.3 from $\Omega = 3 \times 10^{-5} $ to $ 4\times 10^{-5} $ rad.s$^{-1}$ in midMS. The theoretical profiles do not match with that of the simulation at the surface. This is because the radial velocity is forced to zero at the surface in our numerical simulation causing a drop in the amplitude of the simulation profile near the surface.}}
    
     { {On the other hand we see a difference of $\sim$ $10$ in ZAMS and $\sim$ $12$ in midMS between the theoretical mixing profiles of $4\times 10^{-5}$ rad.s$^{-1}$ and $1\times 10^{-4}$ rad.s$^{-1}$ compared to a difference of $\sim 2$ seen in the simulation profiles (Fig. \ref{MP}) for both the models. We propose it could be due to the fact that we are comparing the theoretical profiles of a single frequency with the simulation profiles which could have contributions from a number of frequencies. At higher rotation, the frequencies contributing to the mixing profiles could be significantly different from that of the slowly rotating model causing the observed difference. Moreover, earlier studies suggest that the rotation reduces the overshoot depth in stellar models \citep{august_2019, Korre_2021}. This could influence the  radius at which the waves are excited particularly at high rotation. Furthermore, the considered theoretical model of GIWs excitation by pressure fluctuations at convection-radiation interface is a simplified one when compared to those of IGWs excitation by turbulent plumes (e.g. \cite{scht_1993}, \cite{pincon2016}). We attribute these factors to the difference noted between the theoretical and simulation profiles of the fastest rotating model. Overall, we can conclude that the decreased mixing we see in our simulations is due to the effect of rotation on reducing the convective motions and hence, reducing wave excitation and propagation.}}
\section{Conclusions}\label{con}

We studied the influence of rotation on mixing induced by IGWs by considering a 7 M$_{\odot}$ model at ZAMS and midMS. Overall we noted that the mixing decreases with increase in rotation irrespective of the age considered.  { {We attribute this to the influence of rotation on convection and the subsequent wave excitation at the convective-radiative interface and the wave propagation in the radiation zone.}} % This decrease can be explained by the decreased radial wavelength and the increased thermal damping experienced by  waves with rotation. We also find that the difference in the mixing between different rotation rates is higher for the midMS models than that of the ZAMS models. We attribute this difference to the presence of Brunt-V\"{a}is\"{a}l\"{a} frequency peak in the older stars where the waves experience an increased thermal damping. We concluded from this observation that the wave generation and propagation is highly influenced by the conditions near the generation region.

 { {The averaged rms velocity in the convection zone decreases with increasing rotation agreeing with the studies of \cite{august_2019, august_2020} particularly for midMS models. We argue that this modification of the convective velocity with rotation is the reason of a less efficient excitation of GIWs and as a consequence of a reduced wave-induced mixing. We noted that the decrease in velocity is higher as we move from $\Omega = 3\times 10^{-5}$ to $4\times 10^{-5}$ rad.s$^{-1}$ in midMS model. We explain this decrease by the increasing critical Rayleigh number with rotation resulting in a dramatically reduced $Re$. This is coherent with the reduction of the rms velocity of the convective mode that carry the more heat predicted by the rotating mixing length theory derived by \cite{steve} and \cite{august_2019} which has been confirmed in local Cartesian high resolution nonlinear simulations computed by \cite{barker_2014}.}} 

%  {We also determined the theoretical diffusion coefficient from the interfacial wave flux adopting the approach of \cite{august_2020}. We find that the theoretical diffusion profiles follow a similar trend as that of the simulation mixing profiles except for the fastest rotating model. We propose the difference noted in the fastest rotating model could be due to the influence of rotation on the overshoot distance at higher rotation.}
%Standard mixing length theory (MLT) \citep{vit} considered in the 1D stellar evolution models neglects the effect of rotation. However a modified approach to the MLT including rotation proposed by \cite{steve} suggests that the rotation inhibits convection. This has been tested and agreed by conducting 3D hydrodynamical simulations \citep{barker_2014}. 
 {While damping in the radiative regions is increased for increasing rotation, %it does not result in a significance difference in the mixing profiles at higher rotation. % 
in real stars, this may be less important than the role rotation plays on reducing convective velocities and hence wave driving.} 
However, we expect the damping to play a significant role near the surface of stars, particularly at later stages of evolution, where the thermal diffusivities are found to be much higher.
Hence, considering the influence of rotation on convection in stars could provide a better constraint to the amplitude with which the IGWs are generated at the interface and the subsequent  mixing caused by these waves in the radiation zone.  {This maybe of great importance for predicting the internal mixing in rapidly rotating intermediate-mass and massive stars \citep{may_nature} and in Pre-Main-Sequence late type stars \citep{corine2013} which are known to rotate faster than our Sun \citep{gallet}}.

 { {The implementation of these mixing profiles in to a 1D stellar evolution code can result in significant changes across ages. Computations of 1D models including time dependant IGWs mixing for different models presented in \cite{varghese_2023} are currently ongoing.  Preliminary results show that inclusion of the mixing by IGWs influences the composition gradient resulting in a more smoothed Brunt-V\"{a}is\"{a}l\"{a} frequency peak compared to the model with a constant envelope mixing that was used (private communications, J. Mombarg, 2024). We expect this modification of the peak by the inclusion of wave mixing to further influence the mixing in the radiation zone of older stars (ongoing work by  Dr. Joey Mombarg and
Dr. May Gade Pedersen).}}

\section{Acknowledgements}\label{ack}
We thank the anonymous reviewer for the helpful comments. We acknowledge support from STFC grant ST/L005549/1 and
NASA grant NNX17AB92G. Computing was carried out on  Rocket High Performance Computing service at Newcastle University and DiRAC Data Intensive service at Leicester, operated by the University of Leicester IT Services, which forms part of the STFC DiRAC HPC Facility (\url{www.dirac.ac.uk}), funded by BEIS capital funding via STFC capital grants ST/K000373/1 and ST/R002363/1 and STFC DiRAC Operations grant ST/R001014/1.
PVFE was supported by the U.S. Department of Energy through the Los Alamos National Laboratory (LANL). LANL is operated by Triad National Security, LLC, for the National Nuclear Security Administration of the U.S. Department of Energy (Contract No. 89233218CNA000001). This work has been assigned a document release number LA-23-32363.  { {SM acknowledges support from the European Research Council through HORIZON ERC SyG Grant 4D-STAR101071505,from the CNES SOHO-GOLF and PLATO grants at CEA-DAp,and from PNPS(CNRS/INSU).}}

\bibliography{sample62}{}

\begin{thebibliography}{}
\expandafter\ifx\csname natexlab\endcsname\relax\def\natexlab#1{#1}\fi
\providecommand{\url}[1]{\href{#1}{#1}}

\bibitem[{{Aerts}(2021)}]{aerts_2021}
{Aerts}, C. 2021, Reviews of Modern Physics, 93, 015001

\bibitem[{{Antoci} {et~al.}(2019){Antoci}, {Cunha}, {Bowman}, {Murphy}, {Kurtz}, {Bedding}, {Borre}, {Christophe}, {Daszy{\'n}ska-Daszkiewicz}, {Fox-Machado}, {Garc{\'\i}a Hern{\'a}ndez}, {Ghasemi}, {Handberg}, {Hansen}, {Hasanzadeh}, {Houdek}, {Johnston}, {Justesen}, {Kahraman Alicavus}, {Kotysz}, {Latham}, {Matthews}, {M{\o}nster}, {Niemczura}, {Paunzen}, {S{\'a}nchez Arias}, {Pigulski}, {Pepper}, {Richey-Yowell}, {Safari}, {Seager}, {Smalley}, {Shutt}, {S{\'o}dor}, {Su{\'a}rez}, {Tkachenko}, {Wu}, {Zwintz}, {Barcel{\'o} Forteza}, {Brunsden}, {Bogn{\'a}r}, {Buzasi}, {Chowdhury}, {De Cat}, {Evans}, {Guo}, {Guzik}, {Jevtic}, {Lampens}, {Lares Martiz}, {Lovekin}, {Li}, {Mirouh}, {Mkrtichian}, {Monteiro}, {Nemec}, {Ouazzani}, {Pascual-Granado}, {Reese}, {Rieutord}, {Rodon}, {Skarka}, {Sowicka}, {Stateva}, {Szab{\'o}}, \& {Weiss}}]{DBdor}
{Antoci}, V., {Cunha}, M.~S., {Bowman}, D.~M., {et~al.} 2019, \mnras, 490, 4040

\bibitem[{{Augustson} \& {Mathis}(2019)}]{august_2019}
{Augustson}, K.~C., \& {Mathis}, S. 2019, \apj, 874, 83

\bibitem[{{Augustson} {et~al.}(2020){Augustson}, {Mathis}, \& {Astoul}}]{august_2020}
{Augustson}, K.~C., {Mathis}, S., \& {Astoul}, A. 2020, \apj, 903, 90

\bibitem[{Baldwin {et~al.}(2001)Baldwin, Gray, Dunkerton, Hamilton, Haynes, Randel, Holton, Alexander, Hirota, Horinouchi, Jones, Kinnersley, Marquardt, Sato, \& Takahashi}]{badwin}
Baldwin, M.~P., Gray, L.~J., Dunkerton, T.~J., {et~al.} 2001, Reviews of Geophysics, 39, 179.
\newblock \url{https://agupubs.onlinelibrary.wiley.com/doi/abs/10.1029/1999RG000073}

\bibitem[{{Ballot} {et~al.}(2010){Ballot}, {Ligni{\`e}res}, {Reese}, \& {Rieutord}}]{ballot2010}
{Ballot}, J., {Ligni{\`e}res}, F., {Reese}, D.~R., \& {Rieutord}, M. 2010, \aap, 518, A30

\bibitem[{{Balona} {et~al.}(1996){Balona}, {B{\"o}hm}, {Foing}, {Ghosh}, {Janot-Pacheco}, {Krisciunas}, {Lagrange}, {Lawson}, {James}, {Baudrand}, {Catala}, {Dreux}, {Felenbok}, \& {Hearnshaw}}]{balona}
{Balona}, L.~A., {B{\"o}hm}, T., {Foing}, B.~H., {et~al.} 1996, \mnras, 281, 1315

\bibitem[{{Barker} {et~al.}(2014){Barker}, {Dempsey}, \& {Lithwick}}]{barker_2014}
{Barker}, A.~J., {Dempsey}, A.~M., \& {Lithwick}, Y. 2014, \apj, 791, 13

\bibitem[{{Belkacem} {et~al.}(2009){Belkacem}, {Samadi}, {Goupil}, {Dupret}, {Brun}, \& {Baudin}}]{belkam_2009}
{Belkacem}, K., {Samadi}, R., {Goupil}, M.~J., {et~al.} 2009, \aap, 494, 191

\bibitem[{{Berthomieu} {et~al.}(1978){Berthomieu}, {Gonczi}, {Graff}, {Provost}, \& {Rocca}}]{berth}
{Berthomieu}, G., {Gonczi}, G., {Graff}, P., {Provost}, J., \& {Rocca}, A. 1978, \aap, 70, 597

\bibitem[{{Braginsky} \& {Roberts}(1995)}]{brag}
{Braginsky}, S.~I., \& {Roberts}, P.~H. 1995, Geophysical and Astrophysical Fluid Dynamics, 79, 1

\bibitem[{{Breton} {et~al.}(2022){Breton}, {Brun}, \& {Garc{\'\i}a}}]{sylvian}
{Breton}, S.~N., {Brun}, A.~S., \& {Garc{\'\i}a}, R.~A. 2022, \aap, 667, A43

\bibitem[{{Brunsden} {et~al.}(2018){Brunsden}, {Pollard}, {Wright}, {De Cat}, \& {Cottrell}}]{dorbruns}
{Brunsden}, E., {Pollard}, K.~R., {Wright}, D.~J., {De Cat}, P., \& {Cottrell}, P.~L. 2018, \mnras, 475, 3813

\bibitem[{{Chaboyer} \& {Zahn}(1992)}]{chab_zan}
{Chaboyer}, B., \& {Zahn}, J.~P. 1992, 253, 173

\bibitem[{{Chandrasekhar}(1961)}]{chandra}
{Chandrasekhar}, S. 1961, {Hydrodynamic and hydromagnetic stability}

\bibitem[{{Charbonnel} {et~al.}(2013){Charbonnel}, {Decressin}, {Amard}, {Palacios}, \& {Talon}}]{corine2013}
{Charbonnel}, C., {Decressin}, T., {Amard}, L., {Palacios}, A., \& {Talon}, S. 2013, \aap, 554, A40

\bibitem[{{Charbonnel} \& {Talon}(2007)}]{CT_2007}
{Charbonnel}, C., \& {Talon}, S. 2007, in American Institute of Physics Conference Series, Vol. 948, Unsolved Problems in Stellar Physics: A Conference in Honor of Douglas Gough, ed. R.~J. {Stancliffe}, G.~{Houdek}, R.~G. {Martin}, \& C.~A. {Tout}, 15--26

\bibitem[{{Cowling}(1941)}]{cow}
{Cowling}, T.~G. 1941, \mnras, 101, 367

\bibitem[{{Dintrans} \& {Rieutord}(2000)}]{Din2000}
{Dintrans}, B., \& {Rieutord}, M. 2000, \aap, 354, 86

\bibitem[{Dintrans {et~al.}(1999)Dintrans, Rieutord, \& Valdettaro}]{drv_1999}
Dintrans, B., Rieutord, M., \& Valdettaro, L. 1999, Journal of Fluid Mechanics, 398, 271–297

\bibitem[{{Gallet} \& {Bouvier}(2015)}]{gallet}
{Gallet}, F., \& {Bouvier}, J. 2015, \aap, 577, A98

\bibitem[{{Garcia Lopez} \& {Spruit}(1991)}]{gs1991}
{Garcia Lopez}, R.~J., \& {Spruit}, H.~C. 1991, \apj, 377, 268

\bibitem[{{Gough} \& {McIntyre}(1998)}]{GM_filter}
{Gough}, D.~O., \& {McIntyre}, M.~E. 1998, 394, 755

\bibitem[{Hurlburt {et~al.}(1986)Hurlburt, Toomre, \& Massaguer}]{hurl}
Hurlburt, N.~E., Toomre, J., \& Massaguer, J.~M. 1986, The Astrophysical Journal, 311, 563

\bibitem[{Korre \& Featherstone(2021)}]{Korre_2021}
Korre, L., \& Featherstone, N.~A. 2021, The Astrophysical Journal, 923, 52.
\newblock \url{https://dx.doi.org/10.3847/1538-4357/ac2dea}

\bibitem[{{Kumar} {et~al.}(1999){Kumar}, {Talon}, \& {Zahn}}]{Kumar}
{Kumar}, P., {Talon}, S., \& {Zahn}, J.-P. 1999, \apj, 520, 859

\bibitem[{Lecoanet \& Quataert(2013)}]{LQ}
Lecoanet, D., \& Quataert, E. 2013, Monthly Notices of the Royal Astronomical Society, 430, 2363.
\newblock \url{https://doi.org/10.1093/mnras/stt055}

\bibitem[{{Lee} \& {Saio}(1997)}]{LS}
{Lee}, U., \& {Saio}, H. 1997, \apj, 491, 839

\bibitem[{{Maeder} \& {Zahn}(1998)}]{MZ_98}
{Maeder}, A., \& {Zahn}, J.-P. 1998, \aap, 334, 1000

\bibitem[{{Mathis}(2009)}]{mathis2009}
{Mathis}, S. 2009, \aap, 506, 811

\bibitem[{{Mathis} {et~al.}(2014){Mathis}, {Neiner}, \& {Tran Minh}}]{mathis2014}
{Mathis}, S., {Neiner}, C., \& {Tran Minh}, N. 2014, \aap, 565, A47

\bibitem[{{Mathis} {et~al.}(2008){Mathis}, {Talon}, {Pantillon}, \& {Zahn}}]{mathis2008}
{Mathis}, S., {Talon}, S., {Pantillon}, F.~P., \& {Zahn}, J.~P. 2008, \solphys, 251, 101

\bibitem[{{Mombarg} {et~al.}(2021){Mombarg}, {Van Reeth}, \& {Aerts}}]{joey}
{Mombarg}, J.~S.~G., {Van Reeth}, T., \& {Aerts}, C. 2021, \aap, 650, A58

\bibitem[{{Montalban}(1994)}]{mon94}
{Montalban}, J. 1994, \aap, 281, 421

\bibitem[{{Montalban} \& {Schatzman}(1996)}]{ms96}
{Montalban}, J., \& {Schatzman}, E. 1996, \aap, 305, 513

\bibitem[{{Montalb{\'a}n} \& {Schatzman}(2000)}]{mons_2000}
{Montalb{\'a}n}, J., \& {Schatzman}, E. 2000, \aap, 354, 943

\bibitem[{Munk \& Wunsch(1998)}]{munk}
Munk, W., \& Wunsch, C. 1998, Deep Sea Research Part I: Oceanographic Research Papers, 45, 1977.
\newblock \url{https://www.sciencedirect.com/science/article/pii/S0967063798000703}

\bibitem[{{Neiner} {et~al.}(2020){Neiner}, {Lee}, {Mathis}, {Saio}, {Lovekin}, \& {Augustson}}]{neiner2020}
{Neiner}, C., {Lee}, U., {Mathis}, S., {et~al.} 2020, \aap, 644, A9

\bibitem[{{Neiner} {et~al.}(2012){Neiner}, {Floquet}, {Samadi}, {Espinosa Lara}, {Fr{\'e}mat}, {Mathis}, {Leroy}, {de Batz}, {Rainer}, {Poretti}, {Mathias}, {Guarro Fl{\'o}}, {Buil}, {Ribeiro}, {Alecian}, {Andrade}, {Briquet}, {Diago}, {Emilio}, {Fabregat}, {Guti{\'e}rrez-Soto}, {Hubert}, {Janot-Pacheco}, {Martayan}, {Semaan}, {Suso}, \& {Zorec}}]{neiner2012}
{Neiner}, C., {Floquet}, M., {Samadi}, R., {et~al.} 2012, \aap, 546, A47

\bibitem[{{Ouazzani} {et~al.}(2019){Ouazzani}, {Marques}, {Goupil}, {Christophe}, {Antoci}, {Salmon}, \& {Ballot}}]{ozz_2019}
{Ouazzani}, R.~M., {Marques}, J.~P., {Goupil}, M.~J., {et~al.} 2019, \aap, 626, A121

\bibitem[{{P{\'a}pics} {et~al.}(2017){P{\'a}pics}, {Tkachenko}, {Van Reeth}, {Aerts}, {Moravveji}, {Van de Sande}, {De Smedt}, {Bloemen}, {Southworth}, {Debosscher}, {Niemczura}, \& {Gameiro}}]{papi}
{P{\'a}pics}, P.~I., {Tkachenko}, A., {Van Reeth}, T., {et~al.} 2017, \aap, 598, A74

\bibitem[{{Paxton} {et~al.}(2011){Paxton}, {Bildsten}, {Dotter}, {Herwig}, {Lesaffre}, \& {Timmes}}]{mesa1}
{Paxton}, B., {Bildsten}, L., {Dotter}, A., {et~al.} 2011, \apjs, 192, 3

\bibitem[{{Paxton} {et~al.}(2013){Paxton}, {Cantiello}, {Arras}, {Bildsten}, {Brown}, {Dotter}, {Mankovich}, {Montgomery}, {Stello}, {Timmes}, \& {Townsend}}]{mesa2}
{Paxton}, B., {Cantiello}, M., {Arras}, P., {et~al.} 2013, \apjs, 208, 4

\bibitem[{{Paxton} {et~al.}(2015){Paxton}, {Marchant}, {Schwab}, {Bauer}, {Bildsten}, {Cantiello}, {Dessart}, {Farmer}, {Hu}, {Langer}, {Townsend}, {Townsley}, \& {Timmes}}]{mesa3}
{Paxton}, B., {Marchant}, P., {Schwab}, J., {et~al.} 2015, \apjs, 220, 15

\bibitem[{{Paxton} {et~al.}(2018){Paxton}, {Schwab}, {Bauer}, {Bildsten}, {Blinnikov}, {Duffell}, {Farmer}, {Goldberg}, {Marchant}, {Sorokina}, {Thoul}, {Townsend}, \& {Timmes}}]{mesa4}
{Paxton}, B., {Schwab}, J., {Bauer}, E.~B., {et~al.} 2018, \apjs, 234, 34

\bibitem[{{Paxton} {et~al.}(2019){Paxton}, {Smolec}, {Schwab}, {Gautschy}, {Bildsten}, {Cantiello}, {Dotter}, {Farmer}, {Goldberg}, {Jermyn}, {Kanbur}, {Marchant}, {Thoul}, {Townsend}, {Wolf}, {Zhang}, \& {Timmes}}]{mesa5}
{Paxton}, B., {Smolec}, R., {Schwab}, J., {et~al.} 2019, \apjs, 243, 10

\bibitem[{Pedersen {et~al.}(2021)Pedersen, Aerts, Pápics, Michielsen, Gebruers, Rogers, Molenberghs, Burssens, Garcia, \& Bowman}]{may_nature}
Pedersen, M.~G., Aerts, C., Pápics, P.~I., {et~al.} 2021, Nature Astronomy, 5, 715–722.
\newblock \url{http://dx.doi.org/10.1038/s41550-021-01351-x}

\bibitem[{{Pin{\c{c}}on} {et~al.}(2016){Pin{\c{c}}on}, {Belkacem}, \& {Goupil}}]{pincon2016}
{Pin{\c{c}}on}, C., {Belkacem}, K., \& {Goupil}, M.~J. 2016, \aap, 588, A122

\bibitem[{Plumley \& Julien(2019)}]{rrb}
Plumley, M., \& Julien, K. 2019, Earth and Space Science, 6, 1580.
\newblock \url{https://agupubs.onlinelibrary.wiley.com/doi/abs/10.1029/2019EA000583}

\bibitem[{{Press}(1981)}]{Press}
{Press}, W.~H. 1981, \apj, 245, 286

\bibitem[{Ratnasingam {et~al.}(2018)Ratnasingam, Edelmann, \& Rogers}]{rathish2019}
Ratnasingam, R.~P., Edelmann, P. V.~F., \& Rogers, T.~M. 2018, Monthly Notices of the Royal Astronomical Society, 482, 5500.
\newblock \url{https://doi.org/10.1093/mnras/sty3086}

\bibitem[{Ratnasingam {et~al.}(2020)Ratnasingam, Edelmann, \& Rogers}]{Rathish2020}
---. 2020, Monthly Notices of the Royal Astronomical Society, 497, 4231.
\newblock \url{https://doi.org/10.1093/mnras/staa2296}

\bibitem[{{Ratnasingam} {et~al.}(2023){Ratnasingam}, {Rogers}, {Chowdhury}, {Handler}, {Vanon}, {Varghese}, \& {Edelmann}}]{rathish_2023}
{Ratnasingam}, R.~P., {Rogers}, T.~M., {Chowdhury}, S., {et~al.} 2023, arXiv e-prints, arXiv:2305.06379

\bibitem[{Rogers \& Glatzmaier(2005)}]{Rogers2005}
Rogers, T.~M., \& Glatzmaier, G.~A. 2005, Monthly Notices of the Royal Astronomical Society, 364, 1135.
\newblock \url{https://doi.org/10.1111/j.1365-2966.2005.09659.x}

\bibitem[{{Rogers} {et~al.}(2013){Rogers}, {Lin}, {McElwaine}, \& {Lau}}]{R2013}
{Rogers}, T.~M., {Lin}, D.~N.~C., {McElwaine}, J.~N., \& {Lau}, H.~H.~B. 2013, \apj, 772, 21

\bibitem[{{Rogers} \& {McElwaine}(2017)}]{Rogers2017}
{Rogers}, T.~M., \& {McElwaine}, J.~N. 2017, \apjl, 848, L1

\bibitem[{{Samadi} {et~al.}(2010){Samadi}, {Belkacem}, {Goupil}, {Dupret}, {Brun}, \& {Noels}}]{samadi_2010}
{Samadi}, R., {Belkacem}, K., {Goupil}, M.~J., {et~al.} 2010, \apss, 328, 253

\bibitem[{{Schatzman}(1993)}]{scht_1993}
{Schatzman}, E. 1993, \aap, 279, 431

\bibitem[{{Stevenson}(1979)}]{steve}
{Stevenson}, D.~J. 1979, Geophysical and Astrophysical Fluid Dynamics, 12, 139

\bibitem[{{Szewczuk} {et~al.}(2021){Szewczuk}, {Walczak}, \& {Daszy{\'n}ska-Daszkiewicz}}]{szew}
{Szewczuk}, W., {Walczak}, P., \& {Daszy{\'n}ska-Daszkiewicz}, J. 2021, \mnras, 503, 5894

\bibitem[{{Takehiro} {et~al.}(2020){Takehiro}, {Brun}, \& {Yamada}}]{tak}
{Takehiro}, S.-i., {Brun}, A.~S., \& {Yamada}, M. 2020, \apj, 893, 83

\bibitem[{Talon \& Charbonnel(2005)}]{TC5}
Talon, S., \& Charbonnel, C. 2005, Astronomy \& Astrophysics, 440, 981–994.
\newblock \url{http://dx.doi.org/10.1051/0004-6361:20053020}

\bibitem[{{Van Reeth} {et~al.}(2016){Van Reeth}, {Tkachenko}, \& {Aerts}}]{timothy2016}
{Van Reeth}, T., {Tkachenko}, A., \& {Aerts}, C. 2016, \aap, 593, A120

\bibitem[{{Van Reeth} {et~al.}(2015){Van Reeth}, {Tkachenko}, {Aerts}, {P{\'a}pics}, {Triana}, {Zwintz}, {Degroote}, {Debosscher}, {Bloemen}, {Schmid}, {De Smedt}, {Fremat}, {Fuentes}, {Homan}, {Hrudkova}, {Karjalainen}, {Lombaert}, {Nemeth}, {{\O}stensen}, {Van De Steene}, {Vos}, {Raskin}, \& {Van Winckel}}]{timothy2015}
{Van Reeth}, T., {Tkachenko}, A., {Aerts}, C., {et~al.} 2015, \apjs, 218, 27

\bibitem[{{Vanon} {et~al.}(2023){Vanon}, {Edelmann}, {Ratnasingam}, {Varghese}, \& {Rogers}}]{vanon_2023}
{Vanon}, R., {Edelmann}, P.~V.~F., {Ratnasingam}, R.~P., {Varghese}, A., \& {Rogers}, T.~M. 2023, \apj, 954, 171

\bibitem[{Varghese {et~al.}(2023)Varghese, Ratnasingam, Vanon, Edelmann, \& Rogers}]{varghese_2023}
Varghese, A., Ratnasingam, R.~P., Vanon, R., Edelmann, P. V.~F., \& Rogers, T.~M. 2023, The Astrophysical Journal, 942, 53.
\newblock \url{https://dx.doi.org/10.3847/1538-4357/aca092}

\bibitem[{{Zahn}(1992)}]{zahn_92}
{Zahn}, J.~P. 1992, \aap, 265, 115

\bibitem[{{Zahn} {et~al.}(1997){Zahn}, {Talon}, \& {Matias}}]{zan97}
{Zahn}, J.~P., {Talon}, S., \& {Matias}, J. 1997, \aap, 322, 320

\end{thebibliography}
\bibliographystyle{aasjournal}
% \begin{thebibliography}{}

% \bibitem[Astropy Collaboration et al.(2013)]{2013A&A...558A..33A} Astropy Collaboration, Robitaille, T.~P., Tollerud, E.~J., et al.\ 2013, \aap, 558, A33 
% \bibitem[Bertin \& Arnouts(1996)]{1996A&AS..117..393B} Bertin, E., \& Arnouts, S.\ 1996, \aaps, 117, 393 
% \bibitem[Corrales(2015)]{2015ApJ...805...23C} Corrales, L.\ 2015, \apj, 805, 23
% \bibitem[Ferland et al.(2013)]{2013RMxAA..49..137F} Ferland, G.~J., Porter, R.~L., van Hoof, P.~A.~M., et al.\ 2013, \rmxaa, 49, 137
% \bibitem[Hanisch \& Biemesderfer(1989)]{1989BAAS...21..780H} Hanisch, R.~J., \& Biemesderfer, C.~D.\ 1989, \baas, 21, 780 
% \bibitem[Lamport(1994)]{lamport94} Lamport, L. 1994, LaTeX: A Document Preparation System, 2nd Edition (Boston, Addison-Wesley Professional)
% \bibitem[Schwarz et al.(2011)]{2011ApJS..197...31S} Schwarz, G.~J., Ness, J.-U., Osborne, J.~P., et al.\ 2011, \apjs, 197, 31  
% \bibitem[Vogt et al.(2014)]{2014ApJ...793..127V} Vogt, F.~P.~A., Dopita, M.~A., Kewley, L.~J., et al.\ 2014, \apj, 793, 127  

% \end{thebibliography}



\end{document}